\documentclass[review, numafflabel]{elsarticle}

\usepackage{mathtools}
\usepackage{lineno}
\usepackage{booktabs}
\usepackage{multirow}
\usepackage[table, dvipsnames]{xcolor}
\usepackage{graphicx}
\usepackage{graphics}
\usepackage{boldline} 
\usepackage{array}
\usepackage{caption}
\usepackage[export]{adjustbox}
\usepackage[breaklinks = true]{hyperref}
\usepackage{longtable}
\usepackage{tabularray}
\usepackage{makecell}
\usepackage{xcolor}
\usepackage{float}
\usepackage[export]{adjustbox}
\usepackage{geometry}
\newgeometry{top=1in, bottom=1in,right=1in,left=1in}
\usepackage{gensymb}
\usepackage{eurosym}
\usepackage{xurl}
\usepackage{comment}
\usepackage[T1]{fontenc}
\usepackage{setspace}
\usepackage[skip=6pt, indent=10pt]{parskip}

\usepackage[resetlabels,labeled]{multibib}
\newcites{S}{Supplementary References}
\biboptions{numbers,sort&compress,super}

\journal{Elsevier}

\newcommand{\beginsupplement}{%
	\setcounter{table}{0}
	\renewcommand{\thetable}{S\arabic{table}}%
	\setcounter{figure}{0}
	\renewcommand{\thefigure}{S\arabic{figure}}%
	\setcounter{section}{0}
	\renewcommand{\thesection}{Supplemental \arabic{section}}%
	\setcounter{equation}{0}
	\renewcommand{\theequation}{S\arabic{equation}}
	\setcounter{page}{1}
}

\begin{document}

\let\today\relax
\makeatletter
\def\ps@pprintTitle{%
    \let\@oddhead\@empty
    \let\@evenhead\@empty
    \def\@oddfoot{\footnotesize\itshape
         {} \hfill\today}%
    \let\@evenfoot\@oddfoot
    }
\makeatother

\pretolerance=10000
\tolerance=2000 
\emergencystretch=10pt

\begin{frontmatter}

\title{
Identifying demand-side measures that matter most for Europe's decarbonisation
}

\author[1,2]{Parisa Rahdan\corref{cor1}  }
\cortext[cor1]{Lead contact and corresponding author, Email: parra@dtu.dk}
\author[3]{Mirko Sch{\"a}fer}
\author[4]{Ana Bel{\'e}n Crist{\'o}bal L{\'o}pez}
\author[1,2,5]{Marta Victoria}

\affiliation[1]{organization={Department of Mechanical and Production Engineering and iCLIMATE Interdisciplinary Centre for Climate Change, Aarhus University},%Department and Organization
            addressline={}, 
            postcode={8000}, 
            city={Aarhus},
            country={Denmark}}
            
\affiliation[2]{organization={Department of Wind and Energy Systems, Technical University of Denmark},%Department and Organization
            addressline={Elektrovej, 325}, 
            postcode={2800}, 
            city={Copenhagen},
            country={Denmark}}

\affiliation[3]{organization={INATECH, University of Freiburg, Emmy-Noether-Str. 2},%Department and Organization
             postcode={79110}, 
            city={Freiburg},
            country={Germany}} 
            
\affiliation[4]{organization={Instituto de Energía Solar, Universidad Politécnica de Madrid, C/Alan Turing S/N},%Department and Organization
             postcode={28031}, 
            city={Madrid},
            country={Spain}} 

\affiliation[5]{organization={Novo Nordisk Foundation $CO_2$ Research Center},%Department and Organization
            addressline={Gustav Wieds Vej 10}, 
            postcode={8000}, 
            city={Aarhus},
            country={Denmark}}

\begin{abstract}
 
{European countries pursue a miscellany of historic, emerging, and planned initiatives to transform energy demand through behavioural shifts and end-use efficiency improvements. Yet, these efforts often evolve within fragmented national frameworks that overlook the rapidly evolving supply landscape, risking misaligned investments and diminishing public engagement. We address this need for prioritization by employing a high-resolution model of the European energy system under a net-zero constraint to co-optimise demand and supply strategies across seven sectors, mapping system response to demand mechanisms that reflect real-world practices. Our findings confirm large-scale gains from demand reductions in heating and industry, and show carbon capture reliance is most sensitive to demand reduction in aviation and shipping sectors. For heating, reducing demand peak hours yields the greatest system-level benefit, while critical demand curtailment emerges as the most effective strategy for the power sector, and is also substantially helpful in alleviating high electricity prices for consumers. Smaller, and arguably more attainable, flexibility measures also prove consequential: shifting demand to coincide with solar output delivers noticeable system-wide advantages, even when the shift spans as little as two hours.

}

\end{abstract}


\end{frontmatter}

\section*{Introduction}
\label{sec:intro}

The energy transition hinges on the two pillars of supply and demand. While supply-side shifts dominate in scale and visibility, as renewable deployment struggles to keep pace, demand-side strategies can be the only remaining lever that secures Europe’s path to net-zero. Reductions in average or peak demand across different sectors can result from technological shifts (electric vehicles or heat pumps with flexible operation substituting fossil-based transport or heating provision), more efficient use of final energy (using smaller vehicles, enhanced insulation through buildings retrofitting) or through behavioural changes (less commuting due to home office or setting a lower target for indoor thermostats). Technological changes maintain energy service demand while altering the demand for different energy carriers, mostly through electrification. In contrast, the second and third types of alterations reduce the energy service demand. While most modern energy system models represent technological shifts \cite{victoria2019role}, alterations in energy service demand are typically absent or assumed exogenously due to the difficulties of assigning a cost to them \cite{grubler2018low, costa2021decarbonisation, wiese2024key, barrett2022energy, van2018alternative}. This creates a fundamental gap in our ability to understand, and subsequently plan for, which demand management strategies are most consequential for decarbonisation.

To address this gap, we examine four stylised mechanisms to represent potential alterations in energy service demand: (1) constant demand reduction, (2) peak shaving, (3) temporal shifting, and (4) demand curtailment. For all chosen mechanisms, there are recent concrete real-world examples that prove their possibility at scale. The COVID-19 pandemic triggered long-lasting changes in mobility patterns through remote work, and public information campaigns during Europe's gas crisis in 2022 led to a lower heat demand in winter \cite{huebner2023self}. These are both examples of constant demand reduction. Emergency alerts during a Californian heatwave prompted a 4\% demand drop within minutes \cite{caiso_2022}, illustrating peak shaving in practice. Consumers reacting to negative electricity prices \cite{ yang2026shaping} exemplify temporal demand shifting, while incentive-based programs for both households and industrial consumers \cite{williams2023demand, silva2023demand, chantzis2023potential} that signal customers to reduce demand in certain time periods can be an example of demand curtailment. However, as consumers' motivation for reducing demand remains low globally \cite{Creutzig2022}, achieving long-term demand alterations requires targeted policies and investment, for example enabling rail to compete with air travel for mid-distance trips \cite{IEA_behaviour2023}.

An extensive literature exists where potential reductions in service demand are estimated for different sectors, including buildings \cite{vivier2025meeting}, transport \cite{van2025demand}, and other sectors for sufficiency-oriented scenarios in Europe \cite{wiese2024reducing}. Most of the investigated reductions are constant demand reduction, represented by our first stylised mechanism. These pre-computed demand reductions have been integrated into various modelling frameworks and found to be an effective strategy to attain climate change mitigation while easing the urgency for large-scale deployment of negative emissions technologies (NETs) such as Bioenergy with Carbon Capture and Storage (BECCS) \cite{van2018alternative}. Grubler et al. \cite{grubler2018low} showed that a global 1.5\degree C pathway without NETs is achievable by reducing final energy demand by 40\%. Costa et al. \cite{costa2021decarbonisation} described a 2\degree C target without BECCS through ambitious behavioural and technological changes. Wiese et al. \cite{wiese2024key} found a 40\% service demand reduction feasible in Europe, lowering dependence on carbon sequestration, energy imports, and nuclear power. Demand reductions also mitigate land-use change \cite{van2018alternative,costa2021decarbonisation}, decrease material requirements, and ease the economic and social burden of the transition \cite{creutzig2024demand}. Collectively, these findings affirm the untapped potential of demand interventions. However, most of these studies rely on exogenously defined demand reductions and lack the temporal resolution and detailed sector coupling needed to investigate capacity requirements and cross-sector impacts in a co-optimised framework.

There are more recent studies that incorporate demand alterations endogenously, allowing for co-optimisation with supply-side technologies. Using a high-resolution energy system model, Zeyen et al. \cite{zeyen2021mitigating} co-optimised heating supply technologies and building renovations, showing that heating peak demand was more effective at decreasing cost than annual demand reductions. Our second stylised mechanism reflects this peak-average dynamic by shaving peak demand across sectors. Zhu et al. \cite{zhu2020impact} demonstrated how building thermal inertia can shift heating loads and reduce storage needs, while Riepin et al. \cite{riepin2024spatio} showed cost savings from spatial and temporal load shifting in data centres. Our third stylised mechanism enables temporal demand shifting, but instead of prescribing shift characteristics based on real-world phenomena, the model endogenously determines the timing and extent of demand shifting, allowing insights to emerge from system optimisation. Finally, Brown et al. \cite{brown2024price} showed that incorporating modest demand elasticity improves market dynamics and stabilizes prices. Our fourth stylised mechanism represents a one-step demand flexibility curve, allowing demand curtailment when prices exceed a defined threshold.

In this study, we assess the impact of varying energy service demand using a high spatially and temporally resolved model of a net-zero European energy system, which captures technological shifts reducing final demand, as well as flexibility from storage and transmission. In contrast with previous literature, we do not pre-define demand reductions based on assumed sufficiency or behavioural changes. Instead, we implement four stylised mechanisms that can be assimilated to real-world phenomena: the first two apply exogenous constant and peak demand reductions in different sectors, while the third and fourth allow the system to endogenously determine temporal shifting and curtailment of demand.

\section*{Results}

The following sections outline which demand alteration strategies, and in which sectors, bring the most system-wide benefits overall, which strategy should be prioritized for each energy sector, what is the best way to decrease energy prices for consumers, and how carbon capture is affected throughout the set of scenarios.

\begin{figure}[htb]
\renewcommand{\figurename}{Figure}  \includegraphics[width=0.85\textwidth,center]{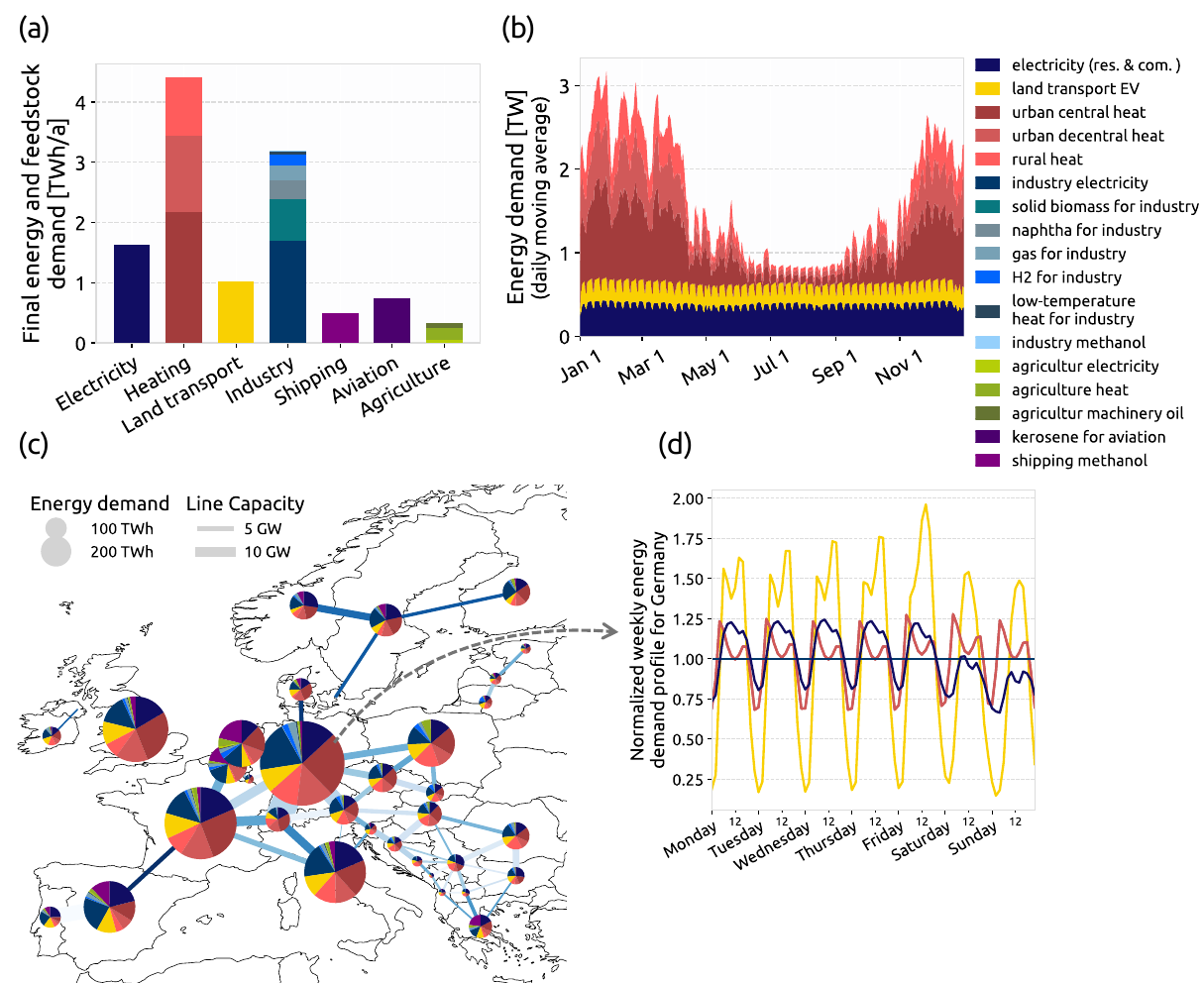}
  \caption{\textbf{Sector demands}. Summary of (a) annual demand per sector, (b) temporal pattern of sectorial demands (rolling daily mean), (c) spatial distribution of demand, and (d) weekly average profile of demand for different sectors included in the model. The sectors are referred to as electricity (residential and commercial), heating (residential and commercial), land transport, industry, shipping, aviation, and agriculture. Only electricity, heating, and land transport have a temporal component. }
  \label{fig:1}
\end{figure}

For all scenarios, capacities and dispatch of technologies are co-optimised for a sector-coupled European energy system using 37 nodes, 2 hourly resolution for one full year, and a net-zero carbon emissions target. The model comprises electricity, heating, land transport, aviation, shipping, industry (including industrial feedstock) and energy consumption in agriculture (Figure \ref{fig:1}). Decarbonising each sector is modelled using both exogenous assumptions (electric vehicles replacing fossil cars) and endogenous decisions (heat pumps, boilers, etc. competing for heat production) that reflect technological transformations for each sector. The model includes several temporal and spatial flexibilities, such as the power transmission grid, batteries, hydrogen, and thermal energy storage. The demand alteration mechanisms are added to the model separately and for different scales in addition to these existing flexibilities. All mechanisms are stylised representations of current demand management strategies, meaning they are implemented with no cost (except demand curtailment). A detailed explanation of all relevant modelling assumptions can be found in the \hyperref[sec:methods]{Methods} section.

\begin{figure}[!htbp]
\renewcommand*{\thefigure}{S\arabic{figure}} \renewcommand{\figurename}{Figure}  
\includegraphics[width=1\textwidth, center]{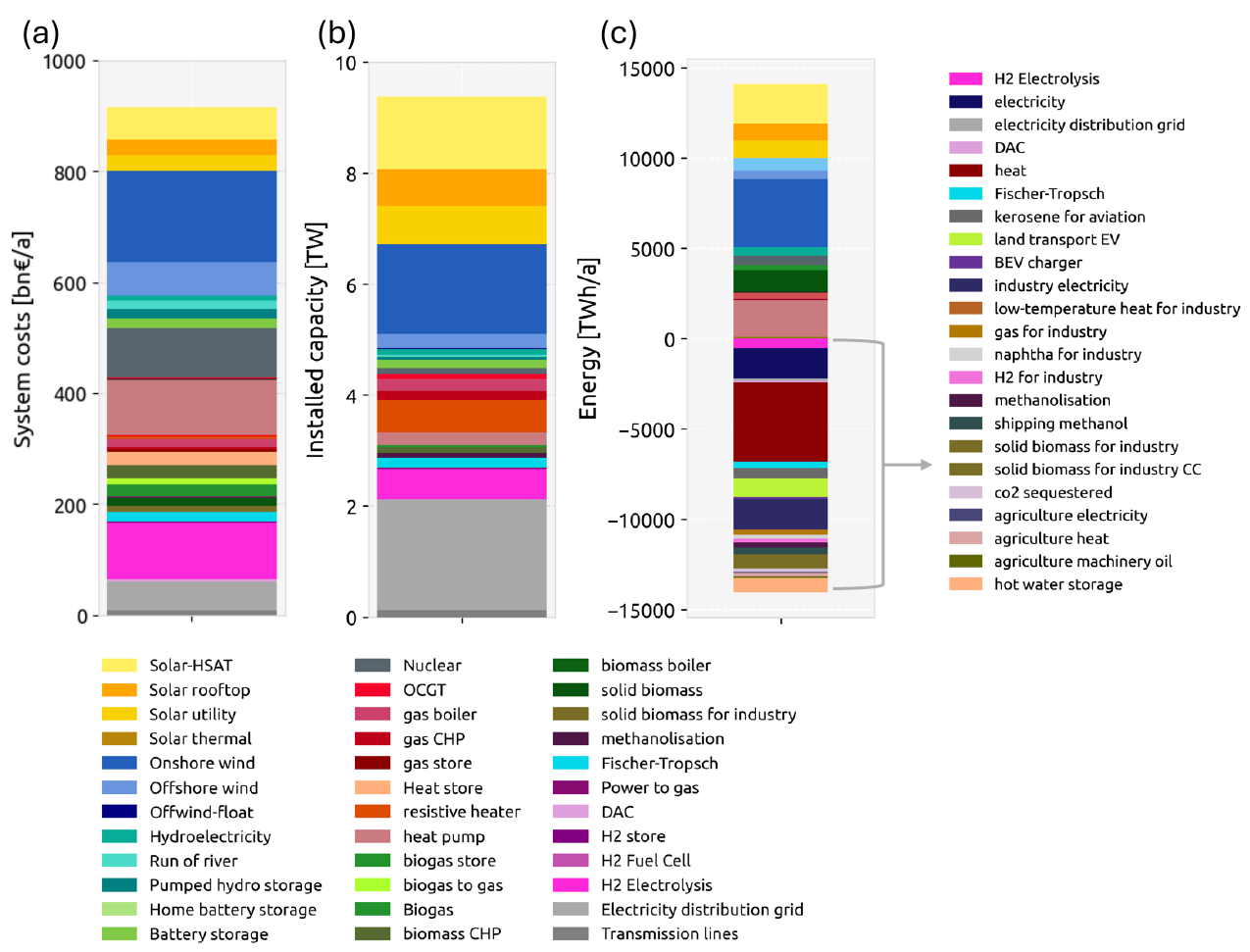}
\caption{\fontsize{9}{10}\selectfont Base scenario composition for (a) total system costs, (b) installed technology capacities, and (c) energy balance. Energy balance shows a summary of all energy generation and energy demands considered in the model.  The system relies primarily on solar PV (utility-scale, rooftop, and horizontal-tracking) and wind (onshore and offshore), which contribute 29\% and 35\% of total generation, respectively. Heat pumps are the next largest energy contributors, followed by biomass, which is used for both electricity generation (CHPs) and heat production in the residential (boilers) and industrial sectors.}
\label{fig:S1}
\end{figure}

The base scenario for a net-zero emissions system carries an annual cost of 936 bn\euro/a and requires a carbon price of 490 \euro/tCO\textsubscript{2} (see \hyperref[sec:methods]{Methods}). The system cost comprises primarily wind and solar capacities, heat pumps and power-to-X technologies (Supplemental information figure \ref{fig:S1}). All service demand alterations in different sectors reduce system cost, albeit depending on the sector and type of demand alteration, the cost reduction and system configuration vary. 

\subsection*{Most effective constant demand reductions for the system}

Our first mechanism, constant demand reduction, represents uniformly lowering demand in a sector, and can be driven by a more efficient use of final demand (improving thermal insulation) or sufficiency-driven (lower thermostat temperature). Our stylised no-cost representation more accurately reflects the sufficiency-driven type, where the service itself is reduced.

For constant demand reduction, system cost decreases linearly with the reduction level for all sectors (Figure \ref{fig:3}a). Heating and industry yield the largest absolute cost savings when demand is uniformly reduced, which is expected as these are the sectors with highest energy demands (Figure \ref{fig:1}a). Reducing their demand cuts the need for costly technologies like heat pumps, electrolysers and Fischer-Tropsch units, and indirectly lowers wind and solar installed capacities. Relative to annual demand however, constant reductions in aviation and shipping demand show the highest benefits (Supplemental information figure \ref{fig:S2}) since these hard-to-abate sectors and their decarbonisation relies on energy-intensive and costly green e-fuels.

Lower demand naturally translates to lower installed generation capacities, with constant demand reductions in heating, industry, and electricity driving the largest decreases in wind and solar deployment. However, certain capacities can increase. Reductions in aviation and shipping emissions enable slightly higher CO\textsubscript{2} emissions in other sectors, which are used primarily by gas boilers. Similarly, lower industrial demand decreases the exogenous demand for biomass which is then used in biomass-fired boilers and CHP units.

\subsection*{Peak shaving is the most effective strategy for heating}

\begin{figure*}%[htb]
\renewcommand{\figurename}{Figure}  \includegraphics[width=0.98\textwidth, center]{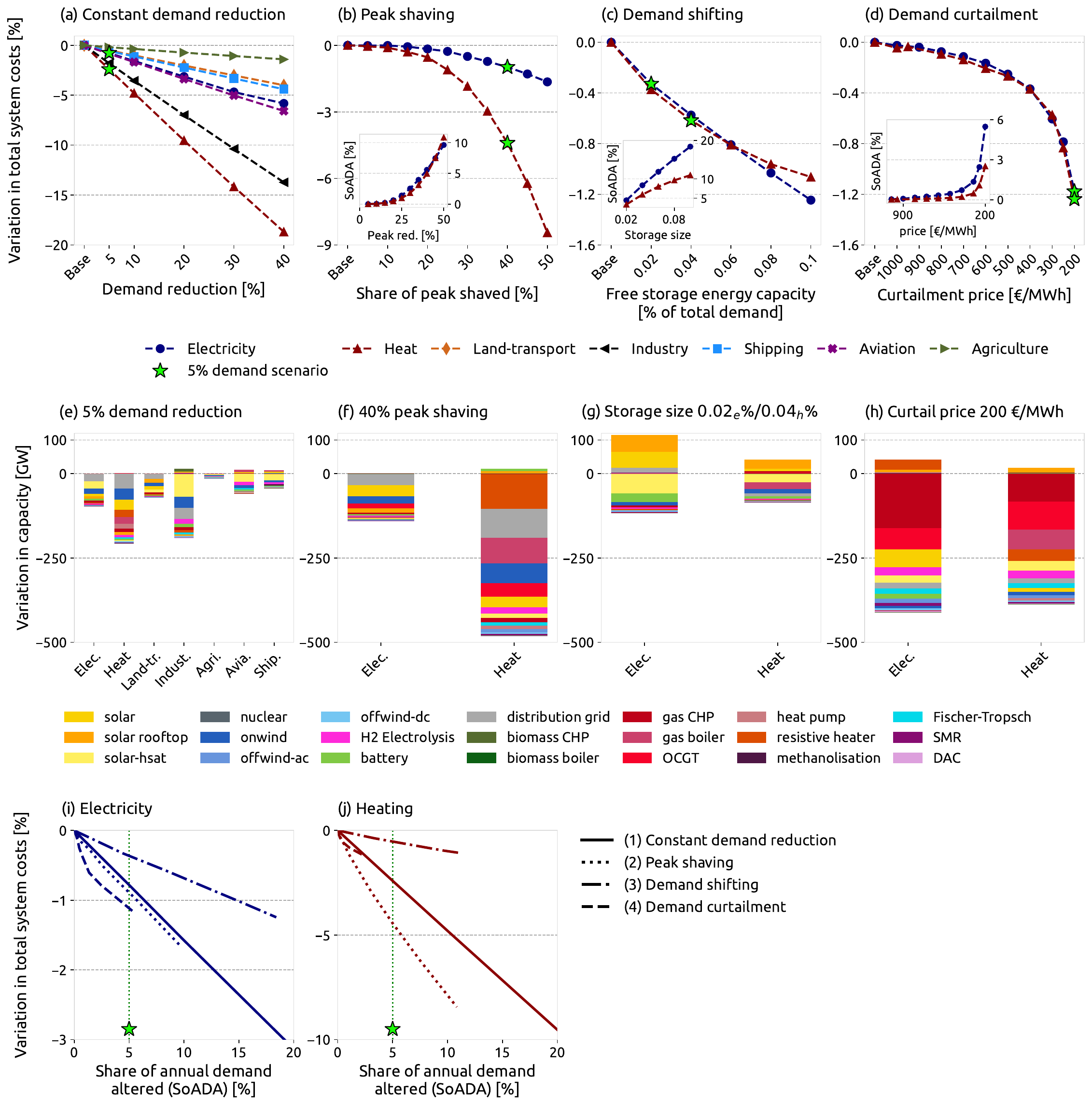}
  \caption{\textbf{System-wide impacts of four demand-alteration mechanisms}. (a,b,c,d) Variation of total system costs for different demand alteration scenarios relative to the base scenario. Scenarios marked with a green star are nearly equivalent in terms of alteration in annual demand being equal to 5\% for electricity or heat sector. Inset figures of (b-d) show share of annual demand altered (SoADA) for each scenario: inset (b) share of annual demand reduction as a function of the peak shaving level, inset (c) share of annual demand shifted for different free storage sizes (size equals ratio of store's energy capacity to annual demand of the sector), and inset (g) share of annual demand curtailed for different load shedding prices (\euro/MWh). (e,f,g,h) Variation of installed capacities of selected system components for green star scenarios relative to the base scenario. (i-j) Comparison of total system cost reduction by the four demand alteration mechanisms for electricity and heating sector (The x-axis shows share of annual demand altered for each scenario). All figures show results for scenarios that reduce demand in one sector (e.g. 40\% demand reduction for aviation in a,e), while demand in other sectors remains the same. }
  \label{fig:3}
\end{figure*}

\begin{figure}[H]
\renewcommand*{\thefigure}{S\arabic{figure}} \renewcommand{\figurename}{Figure}  
\includegraphics[width=0.9\textwidth, center]{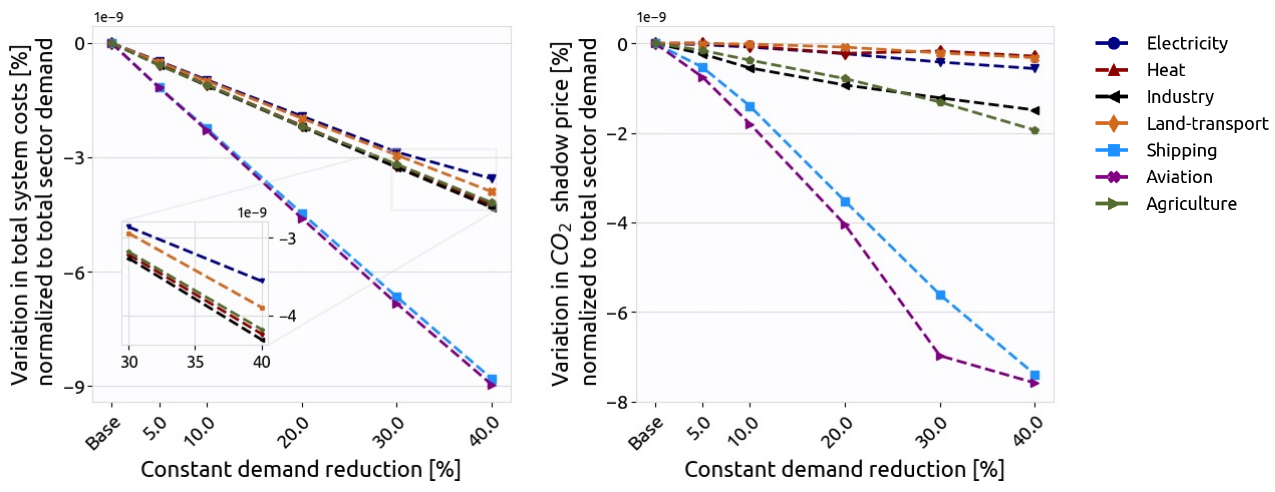}
\caption{\fontsize{9}{10}\selectfont Reduction of:  (a) total system costs and (b) CO\textsubscript{2} price for constant demand reduction in different sectors, normalised to the annual demand of each sector. Aviation and shipping yield the greatest benefits per unit of final energy reduced, making demand reduction in these sectors more effective than in others. However, because their overall demand is relatively low, the absolute savings are smaller (see Figure \ref{fig:3} in the main text). These two sectors depend on high-cost production of synthetic oil (through Fischer-Tropsch units) and methanol, both of which require captured carbon. }
\label{fig:S2}
\end{figure}

In the second set of scenarios, peak shaving is applied to the electricity and heating sectors, which include hourly demand profiles. This mechanism represents consumers lowering their consumption during peak hours, e.g. due to price signals, without cost or shift to later/earlier time, the latter being captured by our third mechanism. Cost savings grow exponentially with the share of peak shaving across all sectors, though at varying rates (Figure \ref{fig:3}b). The heating sector delivers the largest cost savings from peak shaving. Also, when taking into account the absolute share of demand being reduced (Figure \ref{fig:3}i-j), peak shaving is seen as the best strategy for the heating sector.

Supplying heat demand peaks is challenging, and the system gains from peak shaving by reducing installed capacity of resistive heaters and gas boilers. At 40\% peak shaving, annual demand is curbed by 5\% (inset Figure \ref{fig:3}b), but almost twice the cost savings of a 5\% constant heat demand reduction are achieved (Figure \ref{fig:3}j). Up to 10\% peak shaving, savings are similar across electricity and heat sector, but diverge in favour of heat sector for higher shares, delivering 2-3 times more cost benefits. For electricity sector, peak shaving results in lower installed capacity of open-cycle gas turbines (OCGT), which serve as the main backup technology.

\subsection*{Shifting demand to align with solar is a low-effort and robust strategy}

\begin{figure}[!htbp]
\renewcommand*{\thefigure}{S\arabic{figure}} \renewcommand{\figurename}{Figure}  
\includegraphics[width=0.9\textwidth, center]{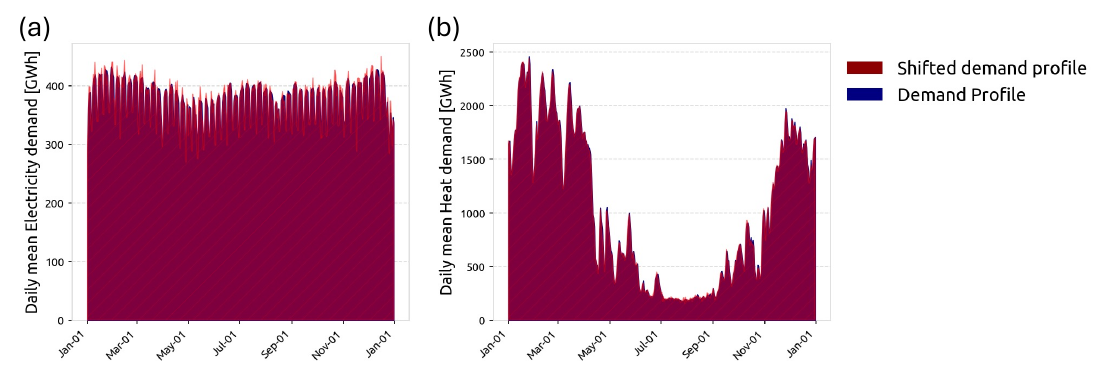}
\caption{\fontsize{9}{10}\selectfont Changes in the rolling daily mean of (a) electricity demand and (b) heating demand when a free storage with size equal to 0.1\% of annual demand for each sector is available. The shifted demand is calculated by adding the net power being discharged from the free storage (negative when charging happens) to the demand. }
\label{fig:S9}
\end{figure}

In the third set of scenarios, we model a free, lossless storage to represent demand shifting via behavioural changes or other means, e.g. using timer-equipped appliances. The energy capacity of the free storage is fixed in each scenario (Figure \ref{fig:3}c). The assumed size for the free storage is limited, enabling only intraday demand shifting rather than seasonal shifts (Supplemental information figure \ref{fig:S9}). Temporal shifting of electricity demand shows a higher impact on system costs than heating because alternative short-term balancing via batteries is more expensive than for heating via thermal energy storage in water tanks. Although the benefits of demand time shifting are smaller than those of the previous two mechanisms for both electricity and heating sector(Figures \ref{fig:3}i-j), it can be argued that it requires less effort, as consumers only need to reschedule their demand rather than reduce it. 

For electricity, a free storage with energy capacity equal to 0.02\% of annual demand (1.8 hours of average demand) enables shifting 4.4\% of total annual demand (Figure \ref{fig:3}c inset). In our analogy, this would require convincing electricity consumers in residential and commercial sectors to relocate around 2 hours of demand to another time of the day. This scenario attains system cost savings of 0.3\% , which is roughly half of those obtained by 40\% electricity peak shaving or constant demand reduction of 5\% (green star scenarios).

\begin{figure}[H]
\renewcommand*{\thefigure}{S\arabic{figure}} \renewcommand{\figurename}{Figure}  

\includegraphics[width=0.9\textwidth, center]{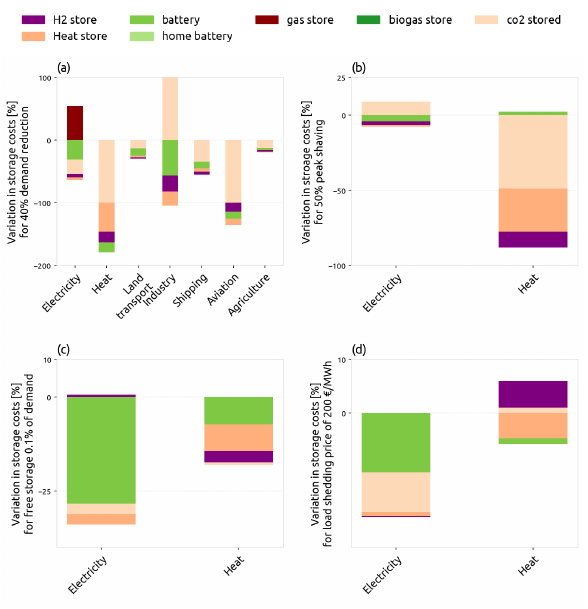}
\caption{\fontsize{9}{10}\selectfont Relative changes in energy capacity of storage technologies for (a) 40\% constant demand reduction, (b) 50\% peak shaving, (c) free storage with size equal to 0.1 \% of annual sector demand, and (d) load shedding with cost of 200 \euro/MWh. }
\label{fig:S6}
\end{figure}

A free storage in the heating sector could mimic the usage of thermal inertia in buildings, e.g. by pre-heating the building during the day and reducing heat demand at night while maintaining indoor comfort levels, as discussed in a previous study \cite{zhu2020impact}. This reduces the need for optimal thermal energy storage capacity (Supplemental information figure \ref{fig:S6}). However, the cost reductions captured by demand-shifting in the heating sectors are limited by (i) the existence of cheap thermal energy storage in the form of water tanks, and (ii) limited energy capacity assumed for the free storage preventing seasonal balancing or peak demand supply, which are the main challenges in this sector.

\begin{figure}[H]
\renewcommand*{\thefigure}{S\arabic{figure}} \renewcommand{\figurename}{Figure}  
\includegraphics[width=0.7\textwidth, center]{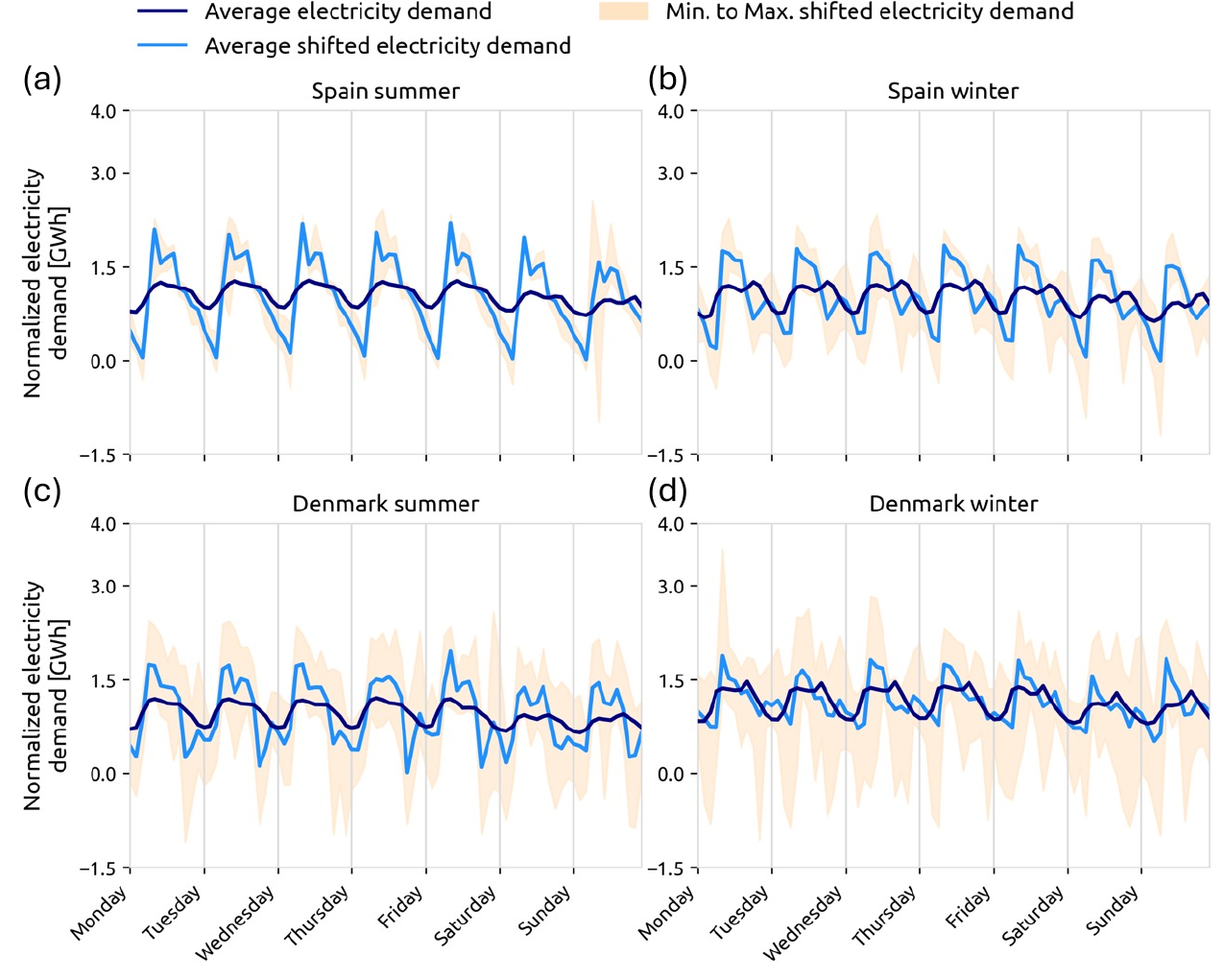}
\caption{\fontsize{9}{10}\selectfont Average weekly profile of electricity demand and shifted electricity demand when a free storage with total size equal to 0.1\% of electricity demand is available for (a) Spain in summer, (b) Spain in winter, (c) Denmark in winter, and (d) Denmark in summer. }
\label{fig:S10}
\end{figure}

For both electricity and heating, the system uses the free storage mimicking behavioural changes to make demand profiles follow solar generation (Figure \ref{fig:4}a). For solar-dominated grids, such as Spain, this pattern is robust. For wind-dominated regions and seasons, such as Denmark in winter, the more random wind generation patterns make the shifted demand profiles less stable across different days (Supplemental information figure \ref{fig:S10}). Demand shifting modifies the optimal capacities, increasing static solar PV, while reducing horizontal single-axis tracking PV and battery storage. Batteries help daily balancing and tracking PV increases morning and evening generation, attributes which lose value when demand is shifted to noon. Since the free storage is assumed to be at the low-voltage level, rooftop PV capacity increases more than utility PV to prevent the losses occurring when transporting utility solar generation via distribution grids \cite{rahdan2024distributed}. Distribution grid capacity also increases to accommodate higher demand at midday.

\subsection*{Curtailing demand during critical hours is the best strategy for electricity}

The fourth set of scenarios includes a stylised representation of demand elasticity by enabling demand curtailment in different sectors when energy prices rise above a certain threshold. The threshold is varied across scenarios, and we do not assume any change in willingness to curtail demand by consumers for lower prices. This mechanism is the most effective for electricity, and 2nd best for heating (Figure \ref{fig:3}i-j). For 200 \euro/MWh, total annual curtailed demand is 5.5\% for the electricity sector (Figure \ref{fig:3}d inset), with total system savings of 1.2\%. The supremacy of this mechanism despite being the only one not modelled as cost-free emphasizes the value of demand curtailment during critical periods. 

System savings from demand curtailment are similar for electricity and heating (Figure \ref{fig:3}d). While one would expect that curtailing electricity is more valuable than heat (since heat pumps can convert 1 kWh of electricity into 3 kWh of heat), demand curtailment occurs in periods of high heat demand in which resistive heaters are extensively used, making the curtailment of one unit of electricity or heat demand roughly equivalent. 

\begin{figure}[H]
\renewcommand*{\thefigure}{S\arabic{figure}} \renewcommand{\figurename}{Figure}  
\includegraphics[width=0.9\textwidth, center]{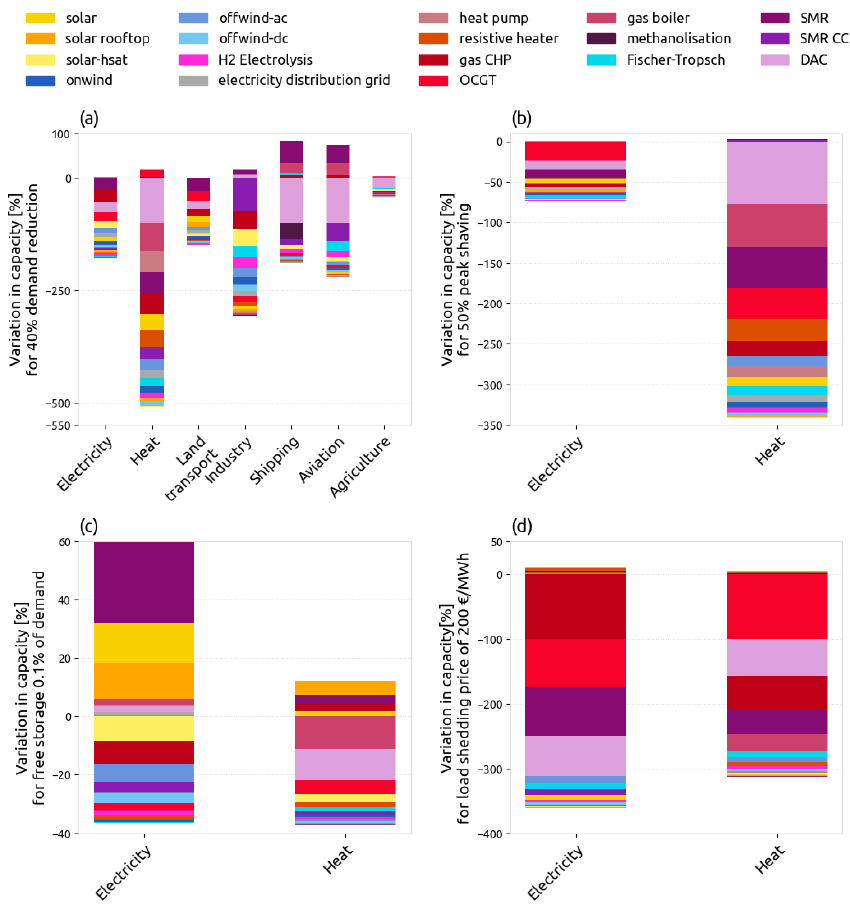}
\caption{\fontsize{9}{10}\selectfont Relative changes in power capacity of different technologies for (a) 40\% constant demand reduction, (b) 50\% peak shaving, (c) free storage with size equal to 0.1 \% of annual sector demand, and (d) load shedding with cost of 200 \euro/MWh. Some technologies such as hydropower, run of river, and pumped hydro storage, are cost-efficient for all scenarios even under high demand reduction and their capacity remains unchanged. }
\label{fig:S4}
\end{figure}

For both electricity and heating sector, allowing demand curtailment reduces dependence on gas turbines and gas-fired CHP plants that are used to cover demand during peak hours with low renewables, and consequently direct air capture (DAC) to compensate for CO\textsubscript{2} emissions of these plants (Figure \ref{fig:3}h and Supplemental information figure \ref{fig:S4}). Heat pumps are sized to cover base heat demand, and their capacity is less affected. When demand is curtailed in electricity sector, the system installs more resistive heaters to meet heat demand.

\begin{figure*}[htb]
\renewcommand{\figurename}{Figure}  \includegraphics[width=0.95\textwidth,center]{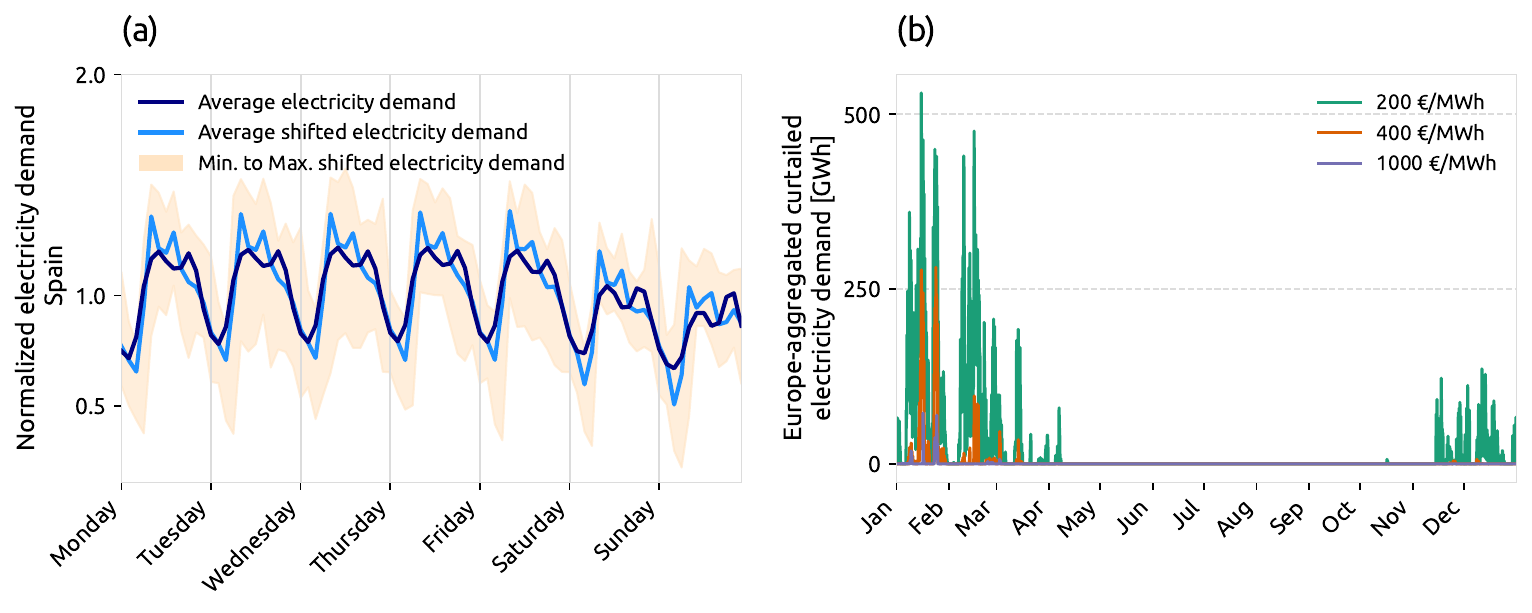}
  \caption{\textbf{Temporal pattern of shifting and curtailing demand} (a) Weekly average electricity demand profile (dark blue) for Spain. The shifted demand (light blue) shows how the profile changes when a free storage with energy capacity equal to 0.02\% of total residential and commercial electricity demand is available. The free storage represents consumer behaviour changes that shift their demand in time. The shifted demand is calculated by adding the net power being discharged from the free storage (negative when charging happens) to the demand. (b) Europe-aggregated curtailed electricity demand time series for different load shedding prices.}
  \label{fig:4}
\end{figure*}

Demand curtailment mainly occurs in winter (Figure \ref{fig:4}b), in periods when renewable generation is low and demand is high. It is also known to be triggered by the overlapping of wind energy droughts, seasonally low solar generation, and high demand driven by low temperatures \cite{gotske2024designing}. Unlike the previous mechanisms, use of demand curtailment is not uniform across Europe (Figure \ref{fig:5}a-d). Germany, Poland, and the Netherlands, which are typical net-importers of energy \cite{rahdan2025strategic}, see the highest demand curtailment relative to annual demand. The value of demand curtailment is highly dependent on the assumed weather year, and naturally the more difficult years with severe energy droughts benefit more from it, but the exact quantification of this relationship is outside the scope of this paper.

\subsection*{Demand mechanisms that benefit consumers the most}

\begin{figure*}[htb]
\renewcommand{\figurename}{Figure}  \includegraphics[width=0.98\textwidth,center]{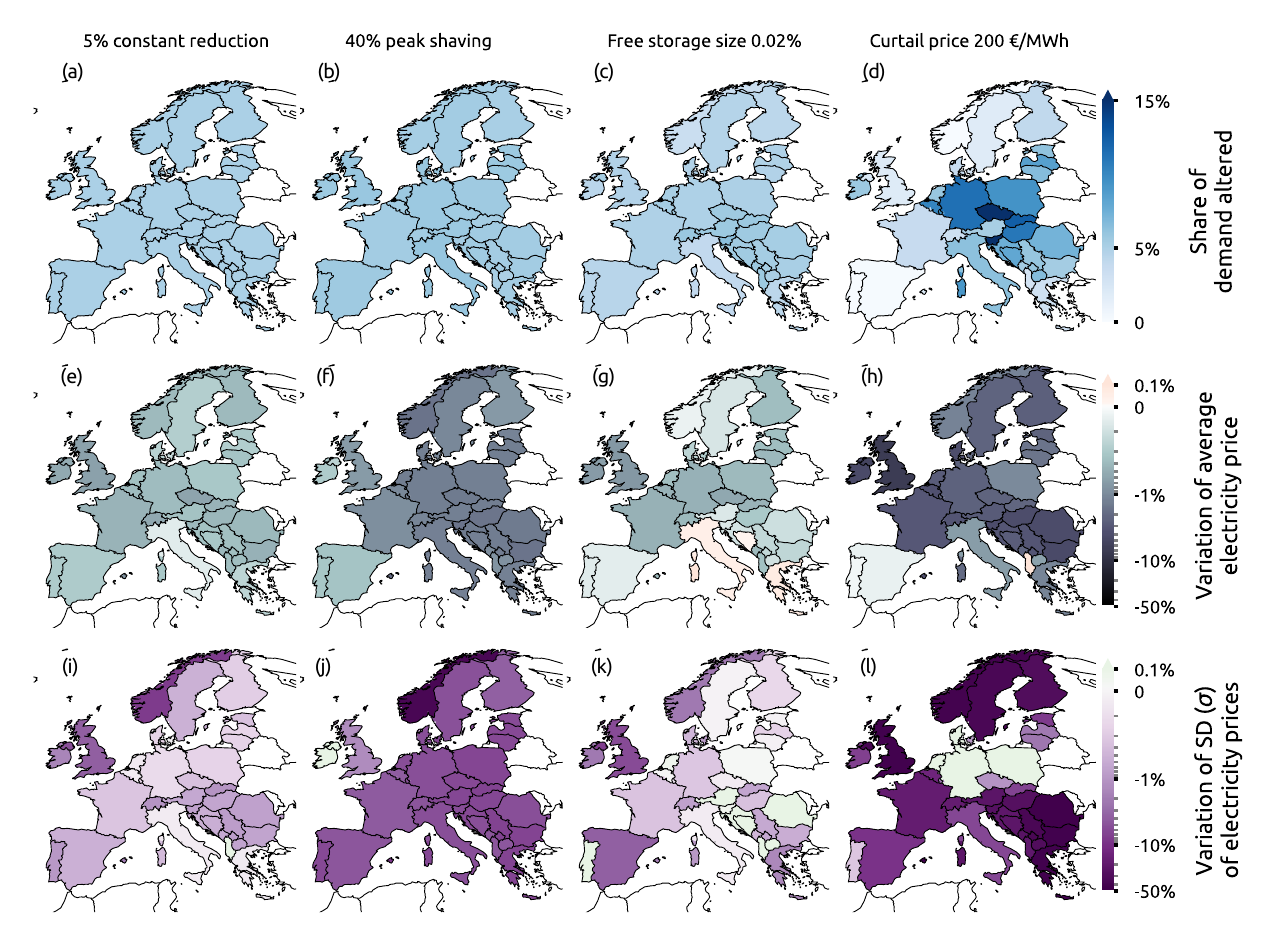}
  \caption{\textbf{Spatial distribution of demand alteration and the benefits} (a-d) Share of annual demand altered, (e-h) variation of average of demand-weighted electricity price, and (i-l) variation of the standard deviation of demand-weighted electricity price for demand alteration scenarios in the electricity sector. The selected scenarios are almost equivalent to 5\% alteration in total electricity demand (they correspond to green star scenarios in Figure \ref{fig:3}).}
  \label{fig:5}
\end{figure*}

\begin{figure}[H]
\renewcommand*{\thefigure}{S\arabic{figure}} \renewcommand{\figurename}{Figure}  
\includegraphics[width=0.55\textwidth, center]{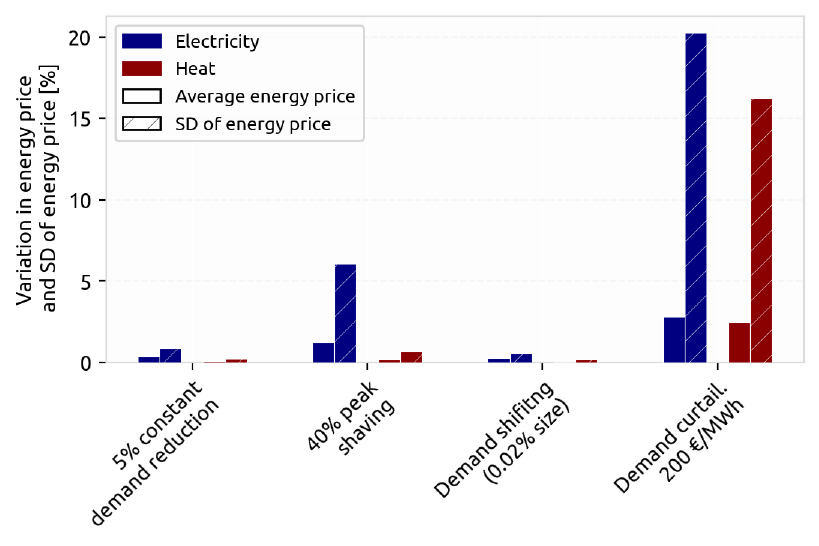}
\caption{\fontsize{9}{10}\selectfont Variation of demand-weighted average price of electricity and heating, and standard deviation (SD) of the prices is shown for green star scenarios of the electricity sector. For example: 5\% constant reduction in electricity demand has a small effect on both average electricity and heat price, while electricity demand curtailment with price of 200 \euro/MWh reduces both prices for consumers. }
\label{fig:S14e}
\end{figure}

It is also relevant to ask what the consumers gain from these demand alterations that benefit the system, as the majority burden of implementing each mechanism is on consumers. Comparing again the four green star scenarios with about 5\% alteration in total demand, peak shaving and demand curtailment are the only mechanisms to noticeably impact electricity average price (demand-weighted) and uniformity for majority of countries in Europe (Figure \ref{fig:5}). The electricity price is based on the shadow price or Lagrange multiplier of the energy balance constraint. Constant demand reduction has negligible impacts on average electricity or heat price and variation (Supplemental information figure \ref{fig:S14e}), because although lower capacities are installed, the difficulty of meeting demand, which sets the price, remains unchanged. A free storage also has negligible impacts for both sectors since it only enables intraday shifting of around 2 hours of demand, providing limited relief during high-price periods driven by renewable energy droughts that can last from days to weeks. 

Consumer benefits are more promising for mechanisms that impact peak demand. However, even 40\% peak shaving reduces prices by only an average of 1.2\% (electricity) and 5\% (heating), with standard deviation dropping by 6\% (electricity) and 12\% (heating). Demand curtailment results in strongest price reduction: 3\% for electricity and 4\% for heating, as well as the largest reduction in price variability: 20\% for electricity and 31\% for heating. This is because demand curtailment allows reducing rarely-used backup capacities and therefore the high energy price hours. 

Both the demand altered and benefits of demand alteration are rather uniform across Europe for constant demand reduction and peak shaving, but there is more spatial variability for the endogenous mechanisms, demand shifting and demand curtailment (Figure \ref{fig:5}). This heterogeneity is due to the system relocating capacities in a way that maximises use of the endogenous mechanism. However, this is mostly an optimisation artifact to minimise total system costs and does not mean demand shifting or curtailment is a disadvantage for consumers in certain countries. The pattern across Europe is therefore not robust for different free storage sizes or curtailment price thresholds. 

It is worth mentioning that consumer benefits from demand alterations in a sector are not restricted to that sector and can spread out across the system. For example, lower electricity prices result in lower electric vehicle charging costs (land-transport sector) and could even reduce heat prices when peak shaving or curtailing electricity demand because backup capacity such as combined heat and power plants is freed up to be used for providing heat supply. Therefore, consumers will gain from electricity demand alteration not just from their electricity bill, but other energy or fuel costs as well.

\subsection*{Best mechanisms for reducing reliance on carbon capture}
\label{chap4:res:co2}

\begin{figure*}[htb]
\renewcommand{\figurename}{Figure}  \includegraphics[width=0.85\textwidth,center]{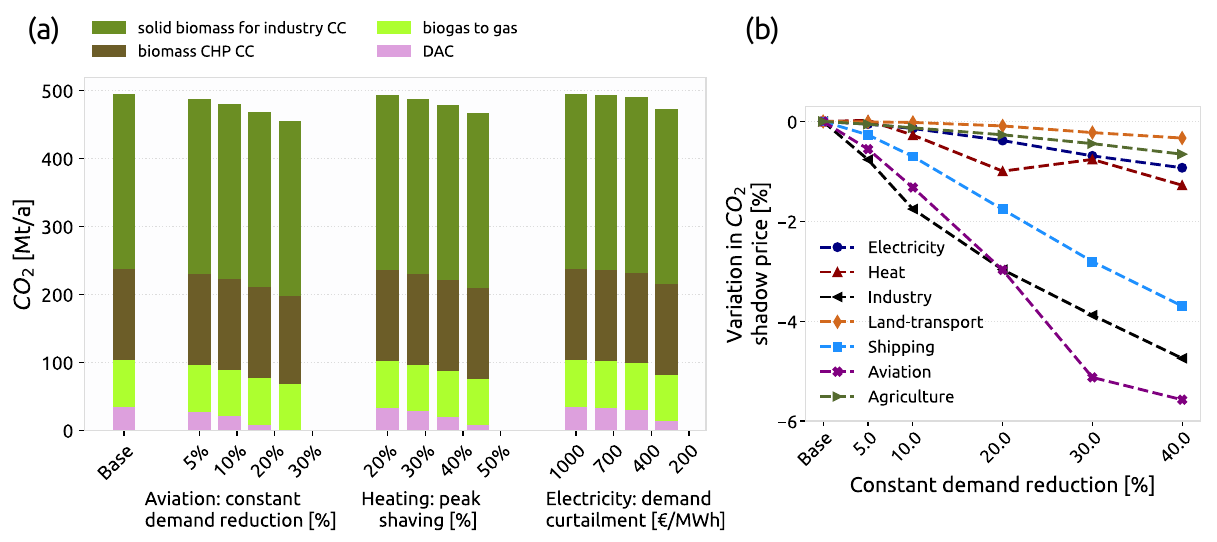}
  \caption{\textbf{Effect of constant demand reduction on CO\textsubscript{2} capture and price}. (a) CO\textsubscript{2} capture in the system for the base and scenarios with significant reduction in direct air capture (DAC). The bar for each scenario shows the different ways CO\textsubscript{2} is captured (from the atmosphere using DAC, point-source biomass, and biogas upgrading. See Supplemental information figure \ref{fig:S19} for more information). (b) Variation of carbon price for constant demand reduction when applied to different sectors. The carbon price indicates the CO\textsubscript{2} tax needed to achieve decarbonisation. 
  }\label{fig:7}
\end{figure*}

Reductions in energy services demand have been suggested as a strategy to attain climate change mitigation with diminished critical reliance on negative emission technologies \cite{grubler2018low, wiese2024key}. Our results show that all demand alteration mechanisms reduce DAC, but only a few can eliminate it, while point-source capture and sequestration are less impacted (Figure \ref{fig:7}a). In addition, the CO\textsubscript{2} price is only reduced noticeably for large constant demand reductions in industry, shipping, and aviation (Figure \ref{fig:7}b).

\begin{figure}[H]
\renewcommand*{\thefigure}{S\arabic{figure}} \renewcommand{\figurename}{Figure}  
\includegraphics[width=0.9\textwidth, center]{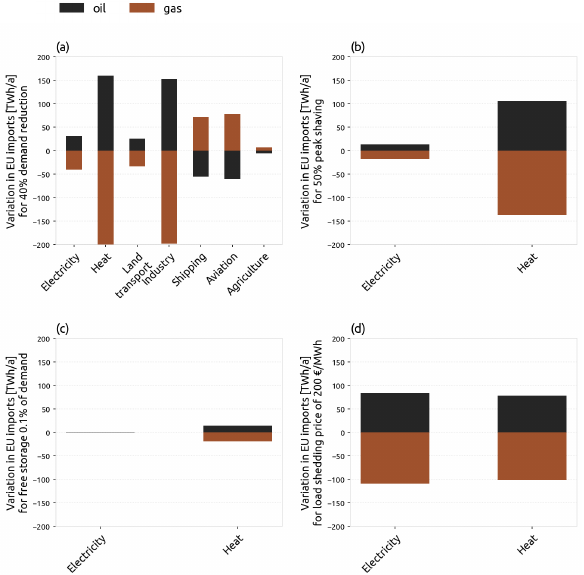}
\caption{\fontsize{9}{10}\selectfont Relative changes in fossil oil and gas imports
 to Europe (TWh/a) for (a) 40\% constant demand reduction, (b) 50\% peak shaving, (c) free storage with size equal to 0.1 \% of annual sector demand, and (d) load shedding with cost of 200 \euro/MWh.}
\label{fig:S7}
\end{figure}

\begin{figure}[!htbp]
\renewcommand*{\thefigure}{S\arabic{figure}} \renewcommand{\figurename}{Figure}  
\includegraphics[width=0.95\textwidth, center]{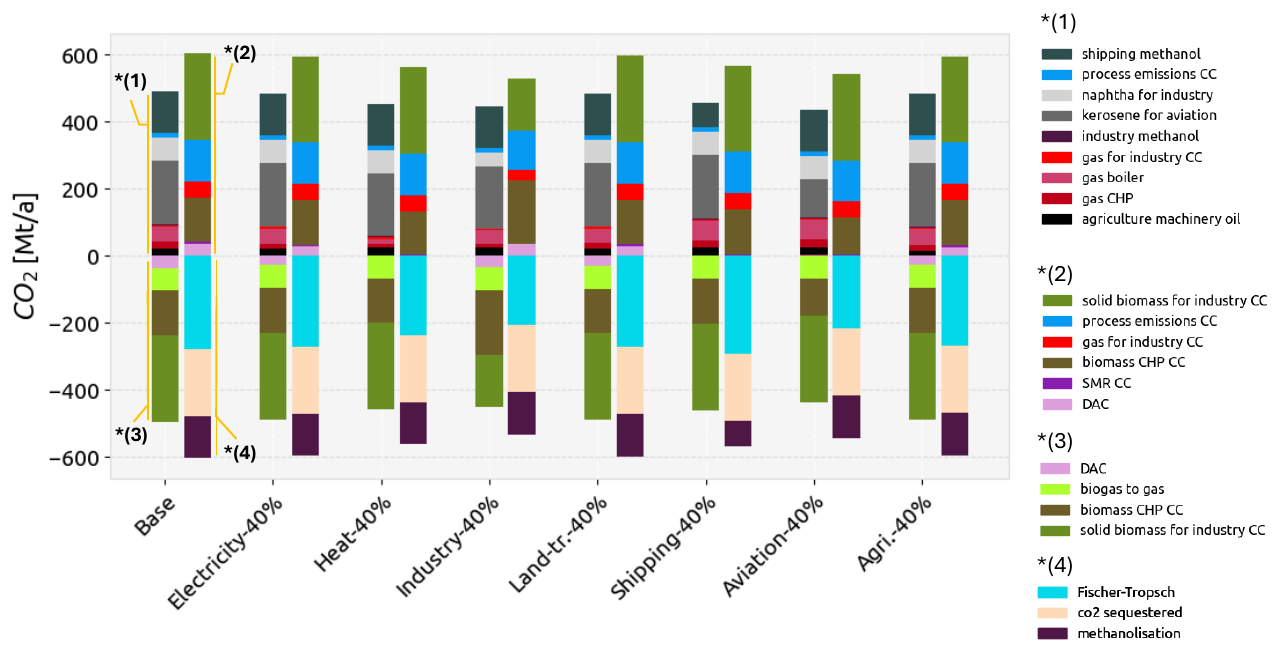}
\caption{\fontsize{9}{10}\selectfont CO\textsubscript{2} balances in the system for the base and 40\% constant demand reduction scenario of different sectors. The left bar for each scenario  shows the CO\textsubscript{2} balance for the atmosphere, i.e. the carbon emitted into the atmosphere (marked *(1)) by technologies such as gas turbines, and the carbon captured (marked *(3)) by technologies such as direct air capture (DAC). The right bar shows the CO\textsubscript{2} balance for the amount of CO\textsubscript{2} stored, with technologies such as biomass CHP with carbon capture storing the CO\textsubscript{2} (marked *(2)), and technologies such as methanolisation utilizing the stored CO\textsubscript{2} (marked *(4)). All scenarios have a maximum carbon sequestration limit of 200 Mega-tonnes/a. The emission intensity of the system (total emitted CO\textsubscript{2} from column *(1) divided by total system demand) rises for electricity-40\% (4\%), heating-40\% (9\%), industry-40\% (2\%), and land-transport-40\% (2\%); and decreases for aviation-40\% (-6\%), shipping-40\% (-10\%), and agriculture-40\% (-1\%).
}
\label{fig:S19}
\end{figure}

To contextualise these results, our model includes a detailed accounting of carbon management where CO\textsubscript{2} can be captured from point sources (biogas upgrading, biomass and gas used in CHP units and industry, process emissions) and DAC, and then be used to produce synthetic carbonaceous fuels (methane, methanol or oil) or sequestered underground, subject to a maximum potential of 200 MtCO\textsubscript{2}/a. This sequestration potential is only slightly higher than the exogenously fixed industrial process emissions in the model (127 MtCO2/a), which limits offsetting fossil emissions and effectively means that almost all exogenous demands for carbonaceous fuels (oil in aviation and industry, methanol in shipping) must be synthetically produced. When service demand reduction in one of the sectors makes attaining carbon neutrality easier, the additional freedom is used to consume more fossil fuels. For instance, constant heat demand reduction raises fossil oil use in aviation, while constant aviation demand reduction increases methane use in heating (Supplemental information figure \ref{fig:S7}). Therefore, although net emissions remain zero across all scenarios, the system's emission intensity per unit of final energy demand rises in some cases (Supplemental information figure \ref{fig:S19}).

Looking at the four demand alteration mechanisms, constant demand reductions of 30\% in aviation or 40\% in heating or shipping eliminate the need for DAC entirely (Figure \ref{fig:7}a); however, this finding is limited to the sectoral emissions represented in our model, which, for instance, excludes non-energy agricultural emissions from fertilizers, livestock, etc. Other mechanisms such as 50\% heat peak shaving or electricity demand curtailment at 200 \euro/MWh price can cut DAC capacity by about 80\%, but can't eliminate it.

\begin{figure}[H]
\renewcommand*{\thefigure}{S\arabic{figure}} \renewcommand{\figurename}{Figure}  
\includegraphics[width=0.95\textwidth, center]{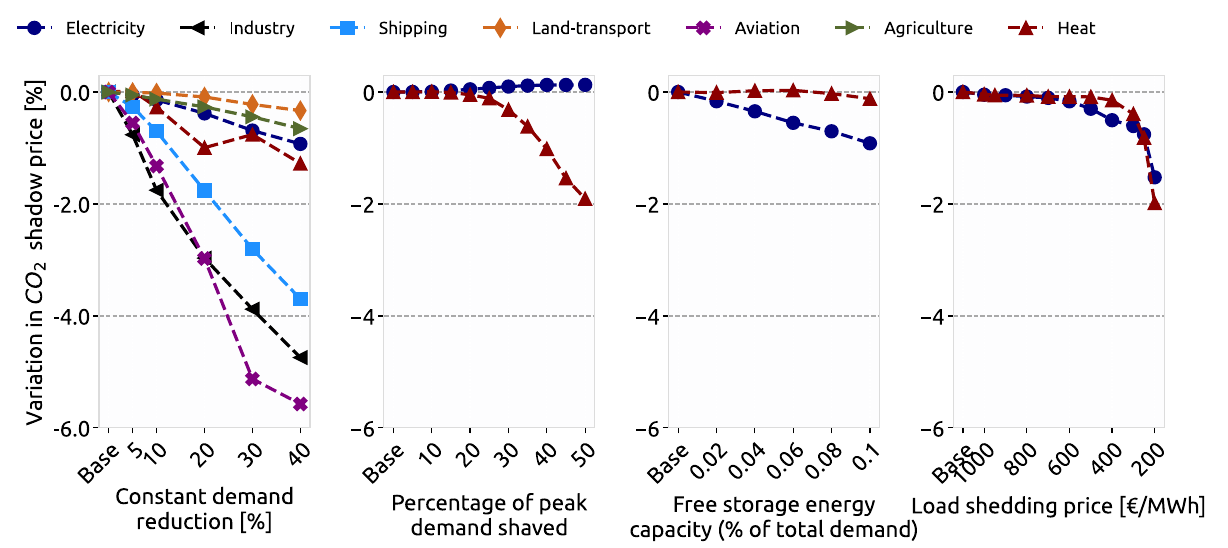}
\caption{\fontsize{9}{10}\selectfont Variation of CO\textsubscript{2} price for different demand alteration scenarios relative to the base scenario. }
\label{fig:S18}
\end{figure}

CO\textsubscript{2} price reductions are also largest for constant reduction in aviation, industry, and shipping (Figure \ref{fig:7}b), reflecting their exogenous demands for oil, gas, and methanol (Figure \ref{fig:1}a). Interestingly, large demand reduction in these hard-to-abate sectors only reduces the CO\textsubscript{2} price by around 5\% while larger variations were obtained when increasing the CO\textsubscript{2} sequestration potential \cite{victoria_2022_S}. For other sectors, all demand mechanisms have a minor impact on CO\textsubscript{2} price (Supplemental information figure \ref{fig:S18}), with heat peak shaving or heat curtailment achieving the largest reductions through lower reliance on gas boilers.

\section*{Discussion}

In this study, we introduce a comprehensive assessment of the economic and environmental benefits of alterations in energy service demand, represented by four stylised mechanisms (constant demand reduction, peak shaving, demand shifting in time, and demand curtailment), in a net-zero sector-coupled European energy system. Our aim is to identify which strategies should be prioritised in each sector, the magnitude of benefits they deliver, and how these benefits scale as greater effort is invested in each mechanism.

We find that constant demand reduction is most effective in the heating sector for lowering total system costs, underscoring the need for more efforts towards building retrofitting \cite{zeyen2021mitigating}. While heating delivers the largest absolute benefits, aviation and shipping achieve the greatest reductions in system costs and carbon prices relative to sector size. These sectors rely heavily on expensive synthetic fuels, and future work could explore more advanced demand-alteration pathways for them, such as electrification. Decreasing demand uniformly does not impact the average price of electricity or heat as generation capacity scales down proportionally while peak demand challenges persist, meaning consumers only benefit through lower energy use from more efficient heating or electrical appliances. 

We have shown from a European system point of view, that the best strategy for heating sector is peak shaving. However, consumer savings are rather low, e.g. they are about 5\% for 40\% heat peak shaving, so it remains arguable whether such savings could motivate consumers to lower their demand during peak hours. For the electricity sector, system benefits from peak shaving and constant demand reduction are on rather equal terms when looking at share of annual demand altered, and consumer savings from peak shaving are even lower since heating peaks play a bigger role in heat price formation than electricity peaks do for electricity price.

Demand shifting in time is particularly beneficial for residential and commercial electricity demand. Here, allowing the system to shift losslessly 2 hours of demand within every day reduces the system cost by 0.3\% (which is already half of the cost reduction attained with a constant 5\% electricity demand reduction). Offering this flexibility to the system makes demand overlap with solar generation. For solar-dominated grids, such as Spain across the year (Figure \ref{fig:4}a) or Denmark in summer, this pattern is robust and suggests the need for public intervention to motivate these demand shifts, including communication campaigns and setting grid tariffs to incentivise consumption when it is sunny. Notice that this is the opposite of historical grid tariffs that incentivised consumption during the nighttime. This should no longer be the case. We do not aim any more for a constant demand profile to reduce the required installed capacity, but for a bell-shaped demand profile to benefit from the cheapest electricity source available \cite{victoria2021solar}.

Demand curtailment was found to be the most effective strategy in the electricity sector for reducing system costs, electricity prices, and price uniformity, but noticeable benefits are only achieved after a certain threshold, e.g. 250 \euro/MWh. Allowing demand curtailment at this threshold reduces total system cost 0.8\% by curtailing 3\% of annual electricity demand, with most of the curtailment happening throughout winter at periods when heat demand is high and renewable generation from wind and solar is low. Although our stylised model might be very optimistic and the compensation might not be high enough to trigger instantaneous reductions in residential and commercial demands, industry consumers have higher demand elasticity (not modelled in our paper). Our results underscore the need for including demand flexibility in cost-optimisation energy system models, as highlighted by Brown and co-authors \cite{brown2024price}, and for quantifying the elasticity of future industry demand (volumes and cost of energy deferral).

A consistent outcome across scenarios is a reduced reliance on direct air capture (DAC), particularly for constant demand reduction in aviation, shipping, and heating, which is arguably de-risking the transition. However, we see that large carbon emission reduction in one sector could lead to higher emissions in another, e.g. using more gas for heat production when aviation demand is reduced. Whether policies that promote demand reductions in one sector while permitting slower decarbonisation in another are practically implementable remains an open question.

Finally, policymakers should note that while renewables are rapidly expanding, energy demand has yet to show a consistent downward trend aligned with the EU’s 2050 targets, calling for targeted action. Policies aimed at behavioural change should be goal-oriented, coordinated with evolving supply-side infrastructure, and set minimum thresholds for demand alterations in order to deliver system-level benefits, as demonstrated in this study. Investments in improving energy intensity and efficiency across sectors should also account for system-wide benefits alongside gains for individual consumers, whether households or industry. Future research could focus on detailed cost-benefit analyses of demand alterations across sectors and mechanisms to support more effective planning toward the 2050 net-zero target.

\section*{Methods}
\label{sec:methods}

\subsection*{Modelling framework}

We use the open sector-coupled model PyPSA-Eur \cite{neumann2023potential} to analyse how changes in energy service demand influence the installed capacity and cost structure of a net-zero European energy system. The model represents Europe using 37 nodes, which are synchronous zones. Simulations are conducted for a full year with a 2-hourly time resolution, assuming a long-term equilibrium in a market with perfect competition and foresight throughout the year, utilizing 2013 weather data and cost assumptions from the technology data v0.9.0 \cite{pypsa_costs} corresponding to 2030.

The optimization problem aims to minimise the total annualised system costs while satisfying critical constraints (see supplementary Note \hyperref[supp:note1]{S1}). We use a greenfield approach, assuming the entire energy system is built from scratch, except for the existing wind, solar, and nuclear installed capacities, existing transmission network of AC and DC lines, and hydropower plants, which are considered exogenously to be at today's capacities. The overnight approach allows for fast optimisation and parallel running of many alternative scenarios, enabling us to focus on the main impacts of different service demand reduction strategies across sectors. Transmission network is based on ENTSOE \cite{ENTSOEtyndp} and can be expanded by 20\% compared to today's volume. The distribution grid is modelled in each node as a single connection from the high-voltage (HV) to the low-voltage (LV) electricity bus with a 97\% efficiency \cite{rahdan2024distributed}.

We set a net-zero CO\textsubscript{2} emissions target. The CO\textsubscript{2} price is obtained as the shadow price or Lagrange multiplier of this constraint, representing the marginal increase in total system cost associated with decreasing the emissions limit by one additional unit. The price is therefore an indicator of the system's difficulty in achieving net-zero emissions and the carbon price required to make certain technologies such as carbon capture economic. A maximum of 200 MtCO\textsubscript{2}/y is allowed to be sequestered underground in every scenario.

\subsection*{Sector demands}

We include seven key sectors in our scenarios (Figure \ref{fig:1}): residential-commercial electricity, residential-commercial heating, land transport, aviation, shipping, industry (including industrial feedstock) and energy demand in agriculture.

The base scenario incorporates technological transformations that modify final energy demands across sectors (e.g. electric vehicles increase electricity demand in road transport and reduce diesel demand), but energy service demands are assumed to remain at today's levels (e.g. passenger-kilometres). Demands for different energy carriers in the base scenario for every sector are shown in Figure \ref{fig:1}. Assumptions for technological transformation follow current European targets. For example, land transport is assumed to be fully electrified by 2050, while aviation remains entirely supplied by liquid fuels, reflecting the absence of mature electrification technologies and the EU's current decarbonisation strategy of using sustainable aviation fuels (SAF) for this sector. Detailed methodologies for constructing demand in every sector taking into account technological transformations have been covered in previous studies \cite{neumann2023potential}, and we only summarize the most relevant information here.

Residential and commercial electricity demand, which we refer to as electricity sector, is represented using hourly demand profiles for each country, derived from 2013 ENTSO-E data \cite{entsoe_load}. These profiles account for electricity use in cooking, cooling, and rail transport amongst others. To allow endogenous optimisation of capacities providing heat, current electrified heating is removed from these profiles. In the base scenario, no exogenous electricity demand growth or reduction is assumed relative to current levels. However, total electricity demand increases endogenously due to e.g. the electrification of heating and the complete transition of land transport to electric vehicles. The primary sources of flexibility in the electricity sector include pumped hydro storage (fixed at today's capacities), as well as endogenously optimised batteries at high-voltage and low-voltage levels, electrolysers along with hydrogen stores, and synthetic production of methane, methanol and oil.

Heat demand encompasses residential and commercial requirements for space heating and hot water, which we refer to as heating sector. Daily profiles for space heating are based on heating degree days and hourly consumption patterns that vary by sector (residential or service), and account for weekdays, weekends, and holidays \cite{IDEES, eurostat}. These profiles are scaled to match each country's annual heat demand in 2013, which is an average year \cite{gotske2024designing}. Spatial distribution of heat demand within countries is weighted by population density, distinguishing between urban and rural areas. 60\% of urban heat demand is assumed to be supplied through district heating systems\cite{zeyen2021mitigating, zhu2020impact}. Thermal energy storage is modelled as small water tanks that allow buffering for a few hours in individual systems and central large water tanks that enable seasonal balancing in district heating systems \cite{victoria2019role}. For the base scenario, we assume space heat demand at today's level, which makes our base results more expensive than other similar net-zero analyses in PyPSA-Eur that by default assume a 29\% space heating reduction due to buildings retrofitting \cite{neumann2023potential}.

Land transport demand is assumed to be fully electrified. Country-level electricity demand for electric vehicles (EVs) is calculated using a fixed EV efficiency of 0.18 kWh/km and the current country-specific energy consumption of internal combustion engine (ICE) vehicles. Therefore, more efficient EVs effectively reduce overall energy demand in this sector in 2050. Hourly electricity demand from EVs is modelled using country-specific driving profiles from the BASt database\cite{bast}, adjusted for local time and incorporating additional heating or cooling loads based on ambient temperature. Only 50\% of idle vehicles allow smart charging and vehicle-to-grid (V2G). Each EV battery is modelled with an average storage capacity of 50 kWh, a charging capacity of 11 kW, and 90\% efficiency for both charging and discharging. EV batteries are requested to be at least 75\% charged by 7 a.m., preventing their use for seasonal storage. The EV fleet and associated infrastructure incur no investment costs in the model, as we assume consumers purchase cars for personal use. 

Aviation demand includes both domestic and international flights, and is supplied by kerosene which can have fossil origin or be synthetically produced via Fischer-Tropsch process. Shipping is assumed to operate entirely on methanol. For both aviation and shipping, annual demands need to be supplied, but the model can decide when to produce the synthetic fuels throughout the year. 

Energy demands in agriculture including electricity, heat, and oil are calculated per country using JRC-IDEES \cite{IDEES} and Eurostat \cite{eurostat} databases, with electricity and heat demands assumed to be constant throughout the year. Similar to aviation, oil for agriculture is an annual demand and can be produced based on endogenous model decision.

The industry sector includes energy and feedstock demands from major European industries such as iron and steel production, chemicals, non-metallic minerals, pulp and paper, and non-ferrous metals. Industrial demands are spatially distributed based on the locations of existing industrial facilities from the Hotmaps Industrial Database\cite{manz2018tf}. While materials production volumes are assumed to remain unchanged from current levels \cite{IDEES}, we implement sector-specific transformations, described in detail by Neumann et al. \cite{neumann2023potential} and summarized here. 
 
Integrated steelworks is fully phased out and substituted by assuming that the recycling route covers 70\% of steel demand and the rest is obtained via direct reduced iron (DRI). Similarly, aluminium recycling is assumed to increase so that it represents 80\% of total demand. Mechanical and chemical recycling of plastics is assumed to cover 30\% and 20\% of total plastic demand. Complete electrification is assumed for low-temperature heat as well as many industrial processes that allow it. Biomass is assumed for medium-temperature heat, while methane and hydrogen are assumed for high-temperature heat. Although certain transformations, such as the share of DRI in steel production, are fixed exogenously, the model incorporates multiple processes to enable endogenous decisions for feedstock production. For instance, hydrogen can be generated via electrolysis or steam methane reforming (SMR) with or without carbon capture. Methane can be produced through the Sabatier process, biogas upgrading, or imported as liquefied natural gas (LNG) from outside Europe. Industrial demands for electricity, low-temperature heat, and hydrogen are assumed to be constant throughout the year. Demands for gas, naphtha, solid biomass, and methanol are imposed annually, and the model can decide when to produce them.

In summary, the model incorporates a range of flexibility mechanisms across sectors to account for temporal and spatial balancing of energy supply and demand. Time-shifting flexibility is provided by storage options for electricity (batteries, PHS, and EV batteries), hydrogen, heat (water tanks), gas, and CO\textsubscript{2} (temporary in overground tanks and permanently when sequestered underground). Additionally, methanol, oil, biogas, and solid biomass can be stored at no cost since only annual balances are imposed. Spatial flexibility is enabled through exchange of electricity (AC and DC transmission lines) and methane (pipelines). Methanol, oil, and solid biomass can also be transported at no cost and without limit in the model. 

\subsection*{Modelling of alterations in energy service demands}

In this section, we describe changes in energy service demands that we investigate in this study, and provide examples of those for different sectors (e.g. passenger-kilometres reduction due to home-office).

\begin{figure}[t]
\renewcommand{\figurename}{Figure} \includegraphics[width=0.5\columnwidth, center]{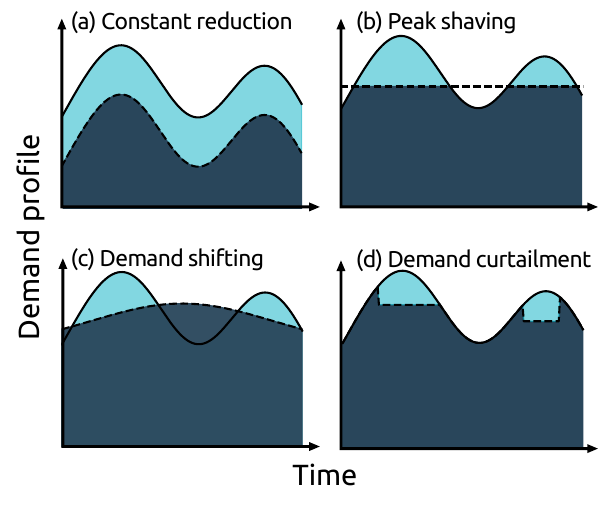}
 \caption{\textbf{Stylised representation of the four demand alteration mechanisms}. (a) Constant reduction (b) Peak shaving (c) Demand time shifting (d) Demand curtailment. }
 \label{fig:2}
\end{figure}

We analyse four stylised representations of service-demand reduction or shifting (Figure \ref{fig:2}). Each of the four demand-side mechanisms is applied independently for different sectors. For example, we apply constant demand reduction separately for each of the seven sectors by 5, 10, 20, 30, and 40\%, while keeping the demand at the reference level in the remaining sectors. This leads to 35 sensitivity scenarios for this mechanism, and we compare a total of 85 scenarios to the base to evaluate the system-wide benefits in terms of total costs, emissions, and installed capacities. We apply service demand alterations at the sector level rather than by energy carrier. For example, reducing demand in the electricity sector (comprising residential and buildings) does not affect electricity use in agriculture or industry, as these are changed under their respective sectors (see Figure \ref{fig:1}a and definitions of the electricity and heating sectors in the previous section). The four service-demand alteration mechanisms are summarized below.

(i)	Constant service-demand reduction: A uniform reduction in demand by a fixed percentage is applied throughout the year (Figure \ref{fig:2}a). Examples of behavioural changes that can be represented by this mechanism include increased use of public transport, cycling, car-sharing, or home-office that reduce passenger-kilometres travelled by EVs, using smaller vehicles or driving them at reduced speeds, setting lower indoor-temperature targets for heating systems, using electrical appliances with higher efficiency in residential or industrial applications, and implementing circular-economy strategies or more conscious consumers that reduce demand for products, and consequently demands in the industry and shipping sectors. Achieving this mechanism in real-life is possible (for example through behaviourally informed interventions \cite{dougru2021best}), but will most likely either entail costs, or be limited to low shares of reduction without compromising human comfort.

(ii) Peak shaving: Under this mechanism, any demand exceeding a certain percentage of the peak demand (defined as the difference between the maximum and minimum demand) is shaved (Figure \ref{fig:2}b). This simulates consumer behaviour changes during peak demand periods, such as lowering their consumption because they react to very high electricity prices published by the market operator in day-ahead markets. In reality, peak shaving is difficult to achieve, and the reduced demand will most likely shift to a later or earlier period \cite{caiso_2022}. A more realistic approach would involve incentives for private and industrial consumers to reduce their consumption, as described below for the demand curtailment mechanism.

(iii) Demand time shifting: This mechanism introduces a cost-free ideal storage at every node, representing consumer behaviour that shifts energy consumption (Figure \ref{fig:2}c). This free storage is located at the same bus as the sector demand, which is the low-voltage (LV) bus for electricity and urban and rural heat buses for heating (see Supplementary Note \hyperref[supp:note1]{S1}). The storage is lossless, unlike the standing loss assumed for hot water storage or the round-trip efficiency of batteries, and there is no limit on the power capacity for discharging or charging it. We vary the energy capacity of the storage, which we refer to as the storage size, from 0.01\% to 0.1\% of annual demand. Examples of this mechanism in the real world are initiatives such as dynamic electricity pricing or time-of-use grid tariffs, which are designed to shift consumers' demand to off-peak periods. Although empirical evidence on the effectiveness of such tariff schemes remains limited, it has been estimated they can reduce peak load by 1-9\% and promote habit formation, suggesting the potential for lasting behavioural changes \cite{enrich2024measuring}.

(iv) Demand curtailment: Instead of considering inelastic demand, this mechanism allows demand curtailment, also known as load shedding, when the price reaches a certain threshold (Figure \ref{fig:2}d). This represents a stylised one-step utility function, i.e., consumers are only willing to consume if price remains below its utility (threshold). More advanced representations of demand flexibility available in PyPSA \cite{brown2024price} are not included here. Examples of this mechanism are interruptibility and demand-response services in Spain and France, where large industrial consumers can be remunerated for making load reductions when requested by the transmission system operator (TSO) \cite{ree2026srad,rte2022reliability}, or demand response programs (DR) in the U.S., where both private and industrial consumers are financially incentivized to reduce consumption during specific hours \cite{williams2023demand}. Data from DR programs indicate the range of peak demand reduced is 0-24\% \cite{nadel2017demandresponse}, while our stylised mechanism has no limit on how much demand can be curtailed at any time-step.

Mechanisms (i) and (ii) are implemented exogenously, meaning demand profiles are altered before optimisation. In contrast, mechanisms (iii) and (iv) allow the model to determine, endogenously, when and where to activate demand shifting or demand curtailment in a cost-optimal way. 

Here, we focus on the system-level impacts of different demand reduction schemes. Potential trade-offs between reductions in system costs and reduced consumer utility or external investment costs to achieve our demand alteration mechanisms are beyond the scope of this study. Therefore, except for demand curtailment, we assume all mechanisms achieved at no cost and do not account for potential rebound effects, such as increased passenger-kilometres travelled due to more efficient vehicles.

\subsection*{Limitations}

We briefly mention here the main limitations of our approach. All our scenarios rely on weather data from a single year, and the optimisation is done with perfect foresight of the full year. Both these assumptions particularly affect the value of demand curtailment during critical periods such as extreme weather events \cite{gotske2024designing}, but quantifying the exact effects is out of the scope of this study. 

All our mechanisms to represent alterations in energy service demands are simplified. In particular, we represent demand elasticity with a one-step function and neglect other, more elaborate, representations \cite{brown2024price}. We also limit modelling of demand curtailment to residential and services electricity demand and heat demand, but do not model the temporal pattern of electricity demand in industry. We do not assess implementation costs of each demand reduction scenario due to uncertainties \cite{zanocco2022assessing, andor2018behavioral, stute2023dynamic}, limiting a complete cost-benefit analysis. Each demand alteration mechanism is examined independently, and the potential interactions between different mechanisms that could augment benefits are not discussed. We also ignore potential sectoral shifts, e.g., remote work lowering transport demand but raising residential electricity and heat demand. 

%\end{linenumbers}

\section*{Data and code availability}
 
The model is implemented using the open energy modelling framework PyPSA and makes use of the model PyPSA-Eur v0.11.0 \cite{pypsa_docs}(available under MIT license via \href{https://github.com/PyPSA/pypsa-eur}{github.com/PyPSA/pypsa-eur/}) and the costs and technology assumptions included in the technology-data v0.9.0 \cite{pypsa_costs}(\href{https://github.com/PyPSA/technology-data}{github.com/PyPSA/technology-data}). Scripts to reproduce the results and figures included in this paper are publicly available at: \href{https://github.com/Parisra/Demand-management-paper}{github.com/Parisra/Demand-management-paper}

\section*{Acknowledgments}
P.R., A.C.L., and M.V. are partially funded by the AURORA project supported by the European Union's Horizon 2020 research and innovation programme under grant agreement No. 101036418. We thank Nikolaj Søltoft Hansen whose M.Sc thesis included a preliminary version of some of the results in this paper.

\subsection*{Declaration of generative AI and AI-assisted technologies in the writing process}

During the preparation of this work the authors used ChatGPT in order to improve readability. After using this tool, the authors reviewed and edited the content as needed and take full responsibility for the content of the published article.

\section*{Author contributions}
P.R. and M.V. jointly conceptualised the study, designed the methodology, contributed to the modelling, and carried out the investigation. P.R. drafted the original manuscript, analysed the data, and created visualisations. M.V., M.S., and A.C.L. contributed substantial reviews of the manuscript. M.V. supervised the investigation.

\section*{Declaration of interests}
The authors declare no competing interests.

\newpage
\addcontentsline{toc}{section}{References}
\bibliography{Bibliography.bib}
%\putbib[Bibliography]
%\end{bibunit}

\clearpage
\onecolumn

\newgeometry{top=25mm}

\addcontentsline{toc}{section}{Supplemental Information}
\beginsupplement
\newgeometry{top=1in, bottom=1in,right=1in,left=1in}
\begin{spacing}{1.15}

\clearpage

\makeatletter
\renewcommand{\theHfigure}{S\arabic{figure}}
\makeatother

\captionsetup[figure]{font={normal}}

\subsection*{Note S1. PyPSA-Eur model}
\label{supp:note1}

PyPSA-Eur models the European energy system by utilizing comprehensive datasets to create demand profiles for sectors including electricity, heating, industry, aviation, shipping, and agriculture. The model also incorporates weather data, renewable energy cost assumptions, land availability for renewable installations, industrial site locations, and the geographical distribution of salt caverns for hydrogen storage (see more details regarding the optimisation constraints, data sources, and technology assumptions in Neumann et al. \protect\cite{neumann2023potential} as well as the model documentation \cite{pypsa_docs}). 

The model encompasses 33 European countries, which include all EU27 members (except Malta and Cyprus), as well as Albania, Great Britain, Montenegro, Norway, Serbia, and Switzerland. The spatial resolution is adjustable, allowing each country to be represented by either a single node or multiple nodes. Spatial aggregation is performed using k-means clustering. Temporal resolution in PyPSA-Eur can range from a single hour to a year. Temporal aggregation is carried out using segmentation clustering via the tsam package \cite{hoffmann2020review}. This approach ensures that variations in supply and demand are preserved by clustering only adjacent snapshots based on their time-series similarity.

The main objective function for the optimisation is to minimise the total annualized system costs by optimizing capacity and dispatch of different technologies for one year, as shown in Eq. (\ref{eq:1})

\renewcommand{\theequation}{S\arabic{equation}}
\begin{equation*}
\min_{G,F,E,P,g,f} = \left[ \sum_{i,r}^{}c_{i,r}.G_{i,r} + \sum_{k}^{}c_{k}.F_{k}+ \sum_{i,s}^{}c_{i,s}.E_{i,s}+ 
\sum_{l}^{}c_{l}.P_{l}+ 
 \right.
\end{equation*}
\begin{equation} \label{eq:1}
\left.\sum_{i,r,t}^{}w_{t}\left(\sum_{i,r}^{}o_{i,r}.g_{i,r,t}+\sum_{k}^{}o_{k}.f_{k,t}  \right)   \right]
\end{equation}
where \(c_{*}\) is capital cost of the component, \(o_{*}\) is operating cost of the component, \(G_{i,r}\) is generator capacity of technology \(r\) at location \(i\), \(E_{i,s}\) is energy capacity of storage \(s\) at location \(i\), \(P_{l}\) is transmission line capacity for line \(l\), \(F_{k}\) is power capacity of technology \(k\) for conversion and transportation of energy, \(g_{i,r,t}\) is generator dispatch of technology \(r\)  at time \(t\), and \(f_{k,t}\) is dispatch of technology \(k\), for instance storage dispatch, or dispatch of transmission lines and gas-to-biogas converters, at time \(t\). Each time snapshot \(t\) is weighted by the time-step \(w_{t}\), and the sum of time-steps is one year. Costs for all technologies and the source for each data are available at the GitHub repository of PyPSA Technology Data \cite{pypsa_costs}.

Various constraints to the optimisation represent different physical and societal limitations in the real-world energy system. For example, demand is usually considered inelastic and must be met at each time-step. A carbon emissions limit is set for a scenario if we want to understand the transition needed to reach climate targets. For each renewable technology, e.g. wind and solar, a maximum renewable potential is calculated in every region based on land availability, and available renewable generation throughout the year is estimated based on weather data. PyPSA-Eur allows customisation of many variables within the model using additional linear constraints, e.g. limiting maximum transmission expansion to reflect social acceptance, or the share of electric vehicles available for use as storage. 

\begin{figure}[htb]
\renewcommand*{\thefigure}{S\arabic{figure}} \renewcommand{\figurename}{Figure}  
\includegraphics[width=0.85\textwidth,center]{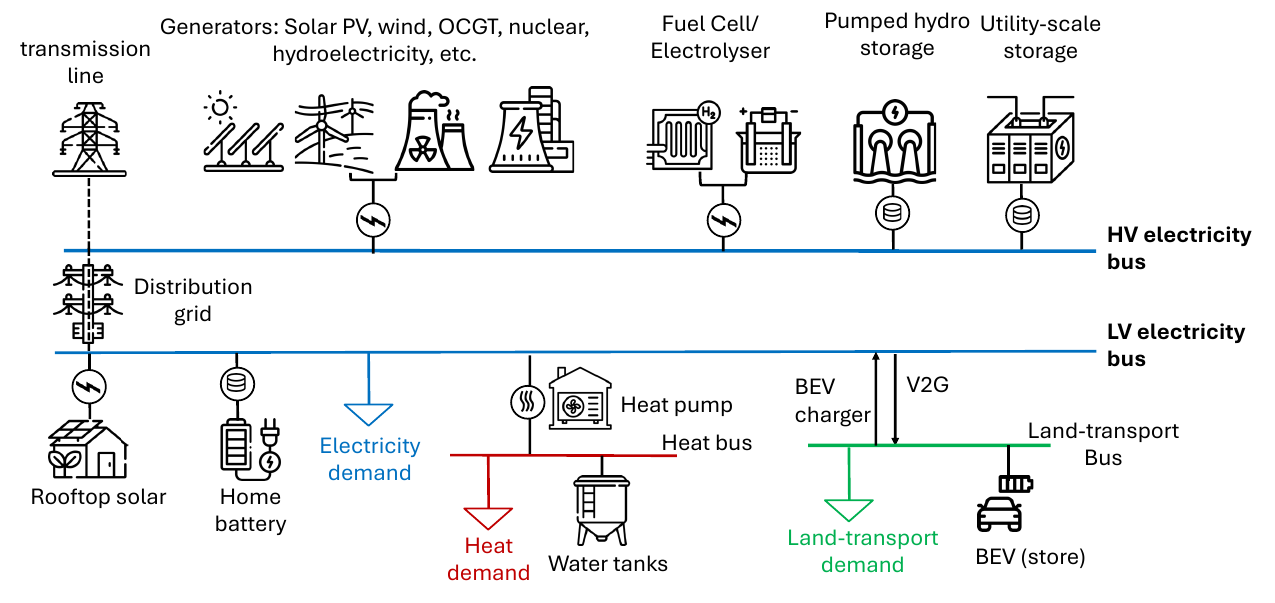}
\caption{\fontsize{10}{14}\selectfont  Simplified representation of a single node when modelling the electricity, heating, and land-transport sector. This figure has been designed using images made by Iconjam (utility battery, heat pump), NeXore88 (Fuel Cell), and freepik (all other icons) from flaticon.com.}
\label{fig:S01}
\end{figure}

\begin{figure}[htb]
\renewcommand*{\thefigure}{S\arabic{figure}} \renewcommand{\figurename}{Figure}   
\includegraphics[width=0.8\textwidth, center]{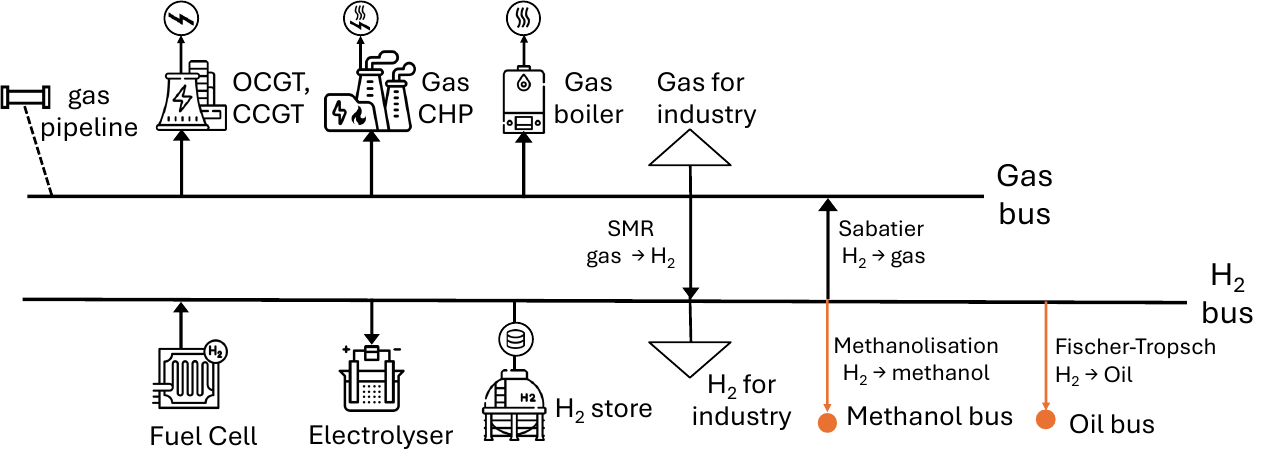}
\caption{\fontsize{9}{10}\selectfont Simplified representation of the gas and hydrogen bus of a single node when modelling all the sectors. For detailed information on different processes within the system refer to Neumann et al. \cite{neumann2023potential}. The land transport bus is where battery electric vehicles (BEV), their chargers (BEV charger), and vehicle-to-grid (V2G) are connected. This figure has been designed using images made by NeXore88 (Fuel Cell) and freepik (all other icons) from flaticon.com.}
\label{fig:S02}
\end{figure}

For each region or node, separate buses that represent different energy carriers are interconnected by links that simulate energy conversions, such as electrolysis (electricity to hydrogen) or fuel cells (hydrogen to electricity). Buses representing the same energy carrier can be connected to each other if they represent a real-world energy transportation method, such as transmission lines between electricity buses. To reduce computational demands, some carriers (e.g., kerosene) are often "copper-plated" and represented with a single European bus, assuming seamless access without transmission costs or losses, which also means ignoring any possible bottlenecks. Figure \ref{fig:S01} shows a simplified representation of how the electricity, heating, and land-transport sector are connected to each other in our model. Figure \ref{fig:S02} shows how the gas bus and the hydrogen bus in each node are connected to electricity and heat generators, storage technologies, pipelines to other nodes, separate demands such as gas for industry, and finally to each other in the form of chemical processes.

\begin{figure}[H]
\renewcommand*{\thefigure}{S\arabic{figure}} \renewcommand{\figurename}{Figure}   
\includegraphics[width=0.6\textwidth,  center]{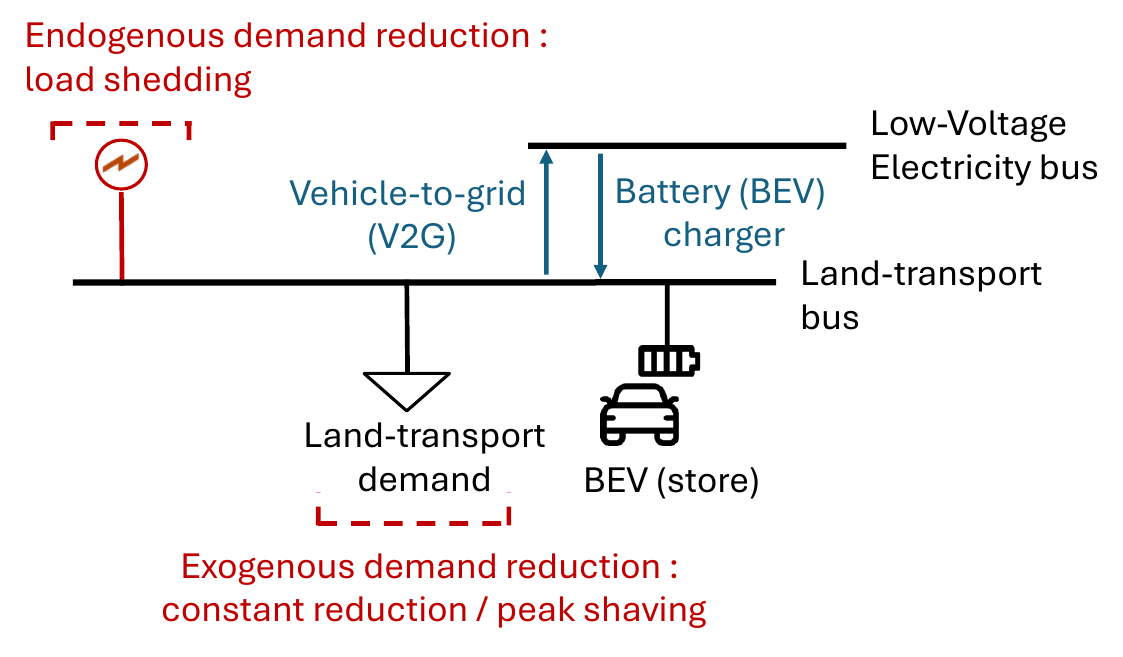}
   \caption{\textbf{} \fontsize{9}{10}\selectfont Representation of how the Land-transport is modeled in PyPSA-Eur, and how different demand reduction scenarios are implemented. The electric vehicle battery (BEV) is modeled as a store with energy capacity of 50 kWh. BEV charging and V2G are modeled as links with 90\% efficiency. Land-transport demand has a daily profile specific to each European country. This figure has been designed using an image made by freepik from flaticon.com}
\label{fig:S03}
\end{figure}

In this study, we include two demand reduction scenarios that represent demand shifting and demand curtailment, the latter involving compensation for private and industrial consumers to reduce their consumption. To model demand shifting, an ideal lossless storage without any charging/discharging constraint is added to the model. For demand curtailment (also called load shedding), a generator with zero capital investment and high marginal cost is added to the model. Both the free storage and the load shedding generator are located at the same bus as the sector demand. For electricity, this is the low-voltage bus. For heating, the free storage and the load shedding generator are added to the urban and rural heat buses, which are connected to the low-voltage bus with technologies such as heat pumps and resistive heaters. For land-transport, demand is located at a separate bus connected to the low-voltage bus at each node, as shown in Figure \ref{fig:S03}.

\clearpage
\section*{Supplementary figures}

\begin{figure}[H]
\renewcommand*{\thefigure}{S\arabic{figure}} \renewcommand{\figurename}{Figure}  
\includegraphics[width=0.95\textwidth, center]{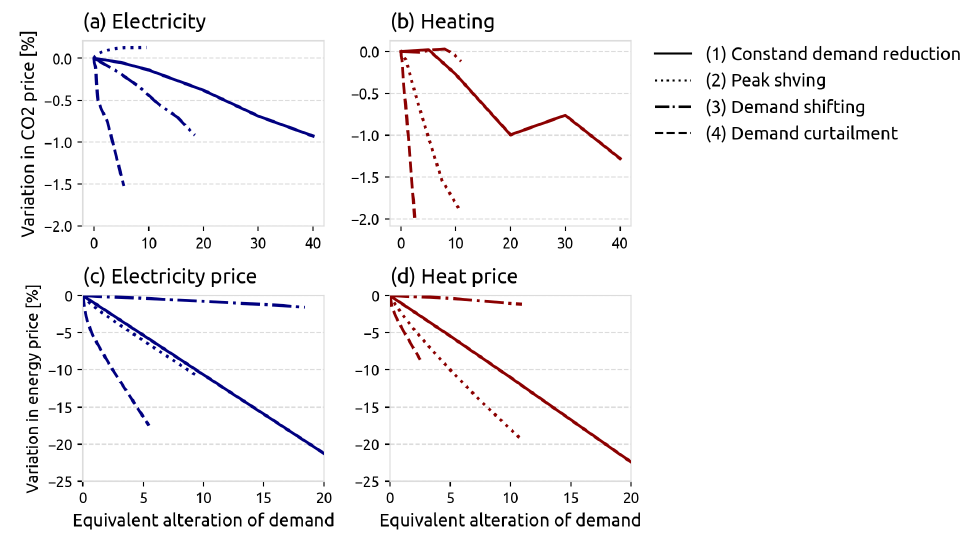}
\caption{\fontsize{9}{10}\selectfont Variation of:  (a,b) CO\textsubscript{2} prices and (c,d) demand-weighted average shadow price of electricity and heat for all four demand alteration mechanisms, compared for electricity and heating sector. The x-axis of each plot represents equivalent annual demand alteration scenarios. For example, shaving electricity peaks by 50\% reduces annual demand by about 10\% (see inset of Figure \ref{fig:3}b in main text), so this scenario is represented by 10\% demand alteration. }
\label{fig:S3}
\end{figure}

\begin{figure}[H]%[!ht]
\renewcommand{\figurename}{Figure}   \includegraphics[width=0.9\textwidth,center]{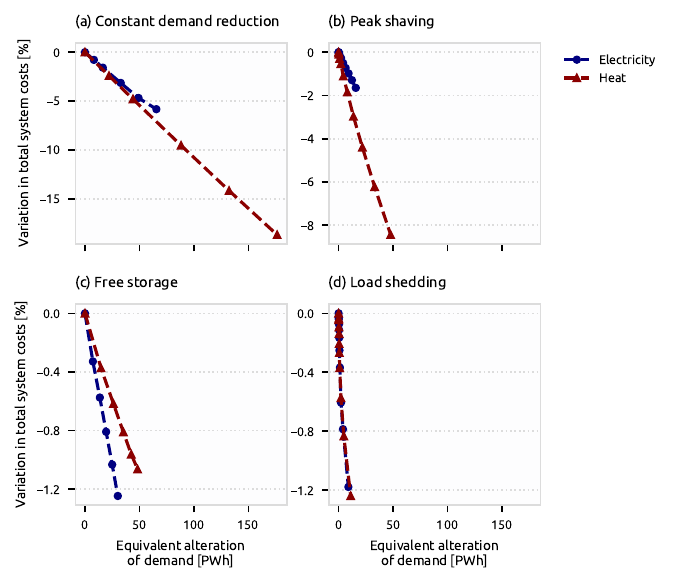}
   \caption{ Variation in total system costs for electricity and heating sector is compared across the four demand alteration mechanisms. The x-axis of each figure shows equivalent alteration of demand: for example, sub-figure (a) shows reducing 1 unit of electricity or heat by constant demand reduction is almost equally beneficial for the system. 
   }\label{fig:S3e}
\end{figure}

\newpage

\newpage

\begin{figure}[H]
\renewcommand*{\thefigure}{S\arabic{figure}} \renewcommand{\figurename}{Figure}  
\includegraphics[width=0.9\textwidth, center]{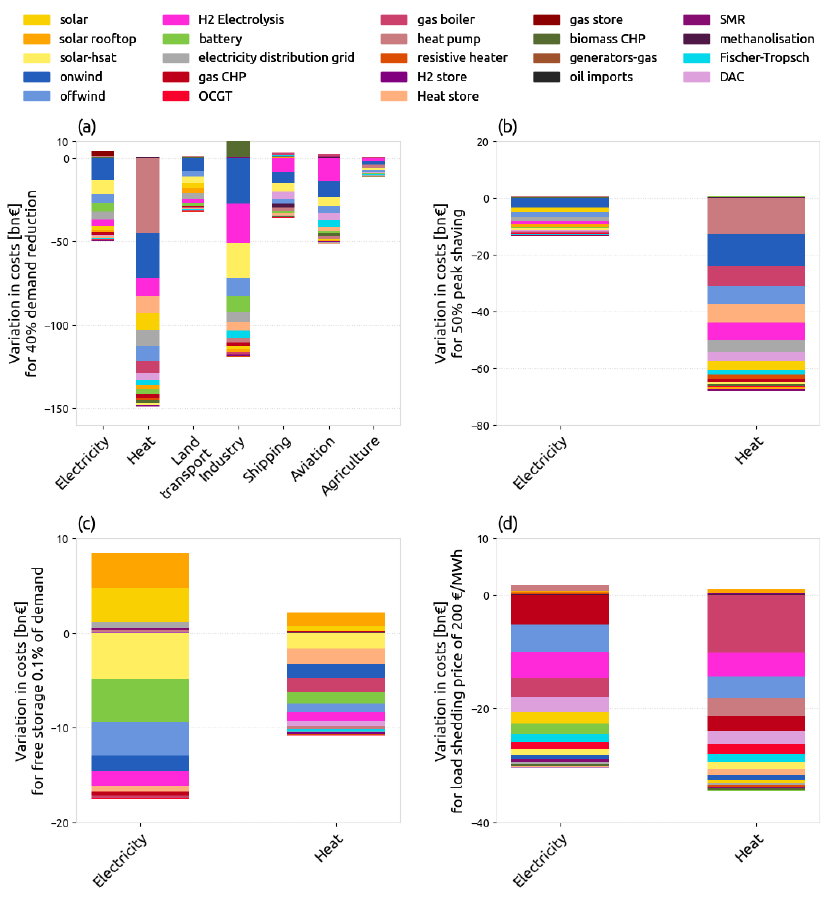}
\caption{\fontsize{9}{10}\selectfont Absolute changes in technology costs for (a) 40\% constant demand reduction, (b) 50\% peak shaving, (c) free storage with size equal to 0.1 \% of annual sector demand, and (d) load shedding with cost of 200 \euro/MWh.}
\label{fig:S5}
\end{figure}

\newpage

\begin{figure}[!htbp]
\renewcommand*{\thefigure}{S\arabic{figure}} \renewcommand{\figurename}{Figure}  
\includegraphics[width=0.85\textwidth, center]{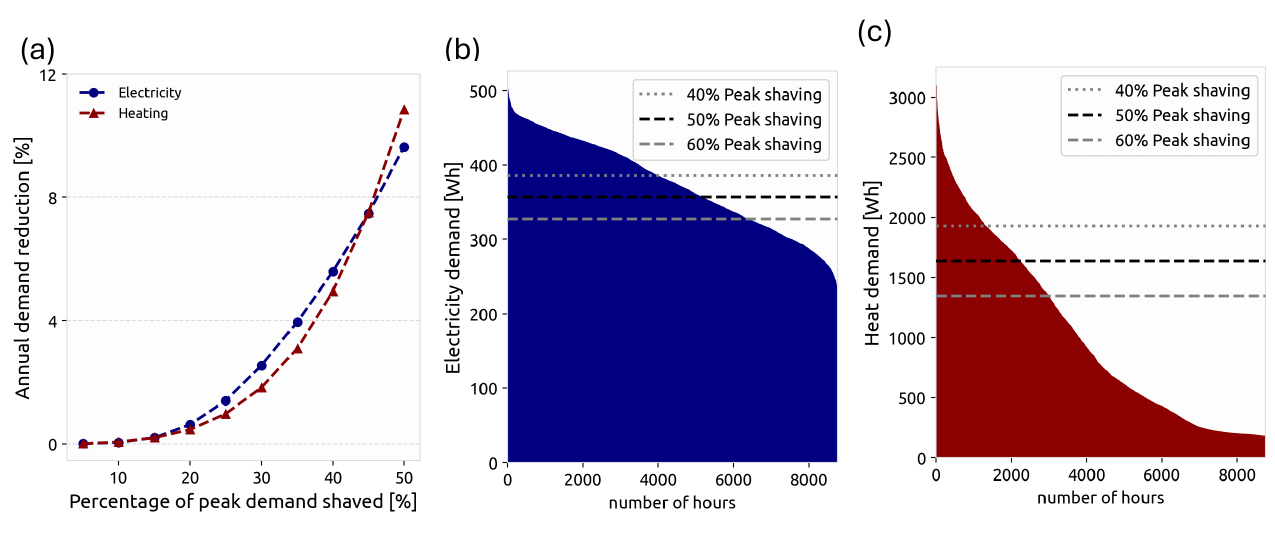}
\caption{\fontsize{9}{10}\selectfont (a) Share of annual demand reduction to peak shaving for electricity, heating, and land-transport. The rest of the figures show duration curves for (b) electricity demand, (c) heating demand, and (d) land-transport demand, and how 40/50/60\% peak shaving would crop the demand curve. Peak shaving is relative to peak demand, which is defined as the difference between the maximum and minimum demand. }
\label{fig:S8}
\end{figure}

\vspace{1.0cm}

\newpage

\begin{figure}[H]
\renewcommand*{\thefigure}{S\arabic{figure}} \renewcommand{\figurename}{Figure}  
\includegraphics[width=0.7\textwidth, center]{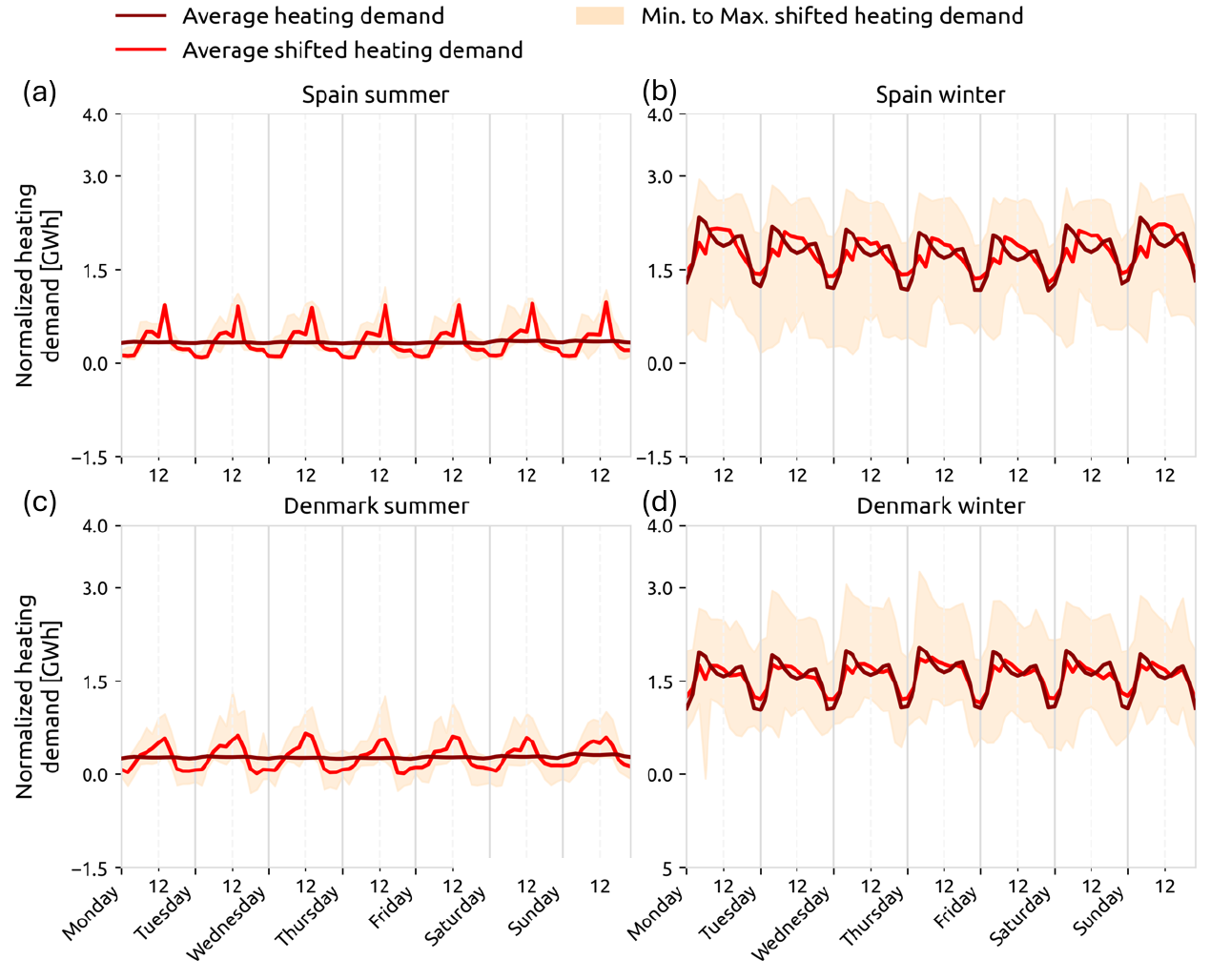}
\caption{\fontsize{9}{10}\selectfont Average weekly profile of heating demand and shifted heating demand when a free storage with total size equal to 0.1\% of heating demand is available at rural and urban heat buses for (a) Spain in summer, (b) Spain in winter, (c) Denmark in winter, and (d) Denmark in summer. }
\label{fig:S11}
\end{figure}

\begin{figure}[H]
\renewcommand*{\thefigure}{S\arabic{figure}} \renewcommand{\figurename}{Figure}  
\includegraphics[width=0.9\textwidth, center]{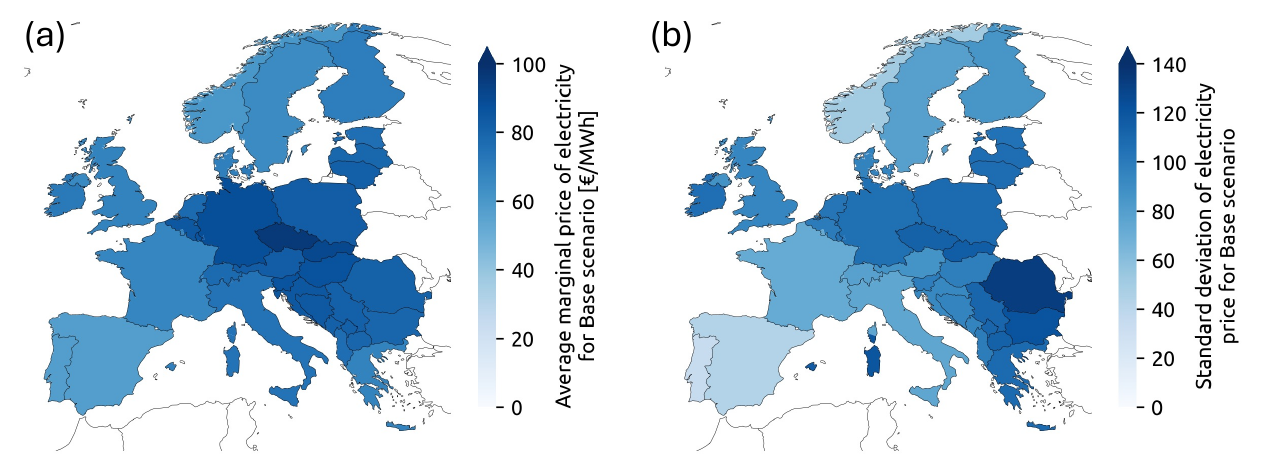}
\caption{\fontsize{9}{10}\selectfont Nodal distribution of (a) average marginal price of electricity and (b) standard deviation of the marginal price of electricity for the base scenario. }
\label{fig:S12}

\newpage

\end{figure}
\begin{figure}[H]
\renewcommand*{\thefigure}{S\arabic{figure}} \renewcommand{\figurename}{Figure}  
\includegraphics[width=0.9\textwidth, center]{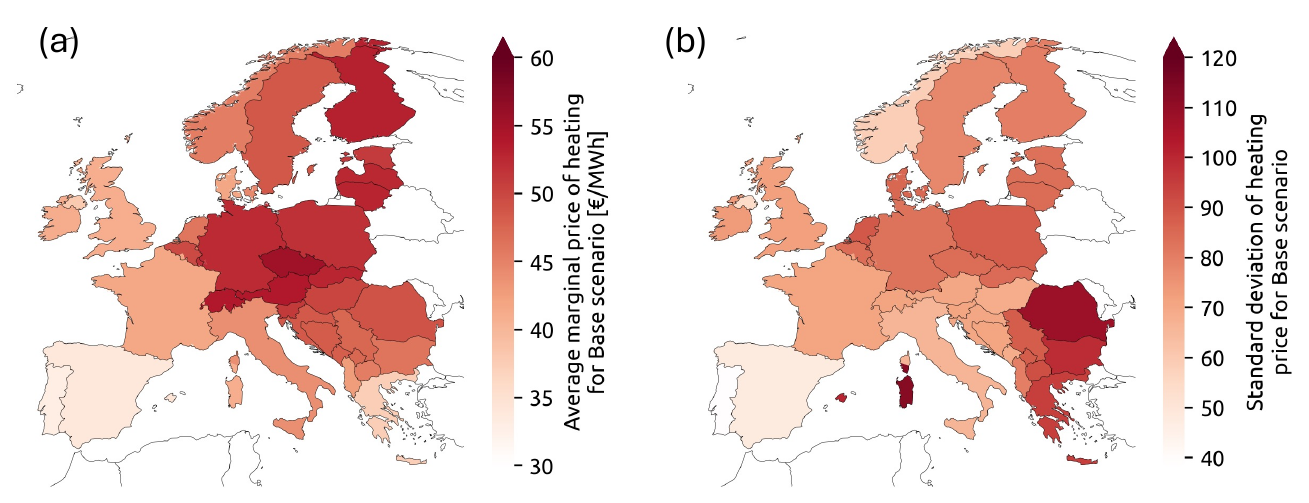}
\caption{\fontsize{9}{10}\selectfont Nodal distribution of (a) average marginal price of heating and (b) standard deviation of the marginal price of heating for the base scenario.}
\label{fig:S13}
\end{figure}

\begin{figure}[H]
\renewcommand*{\thefigure}{S\arabic{figure}} \renewcommand{\figurename}{Figure}  
\includegraphics[width=0.8\textwidth, center]{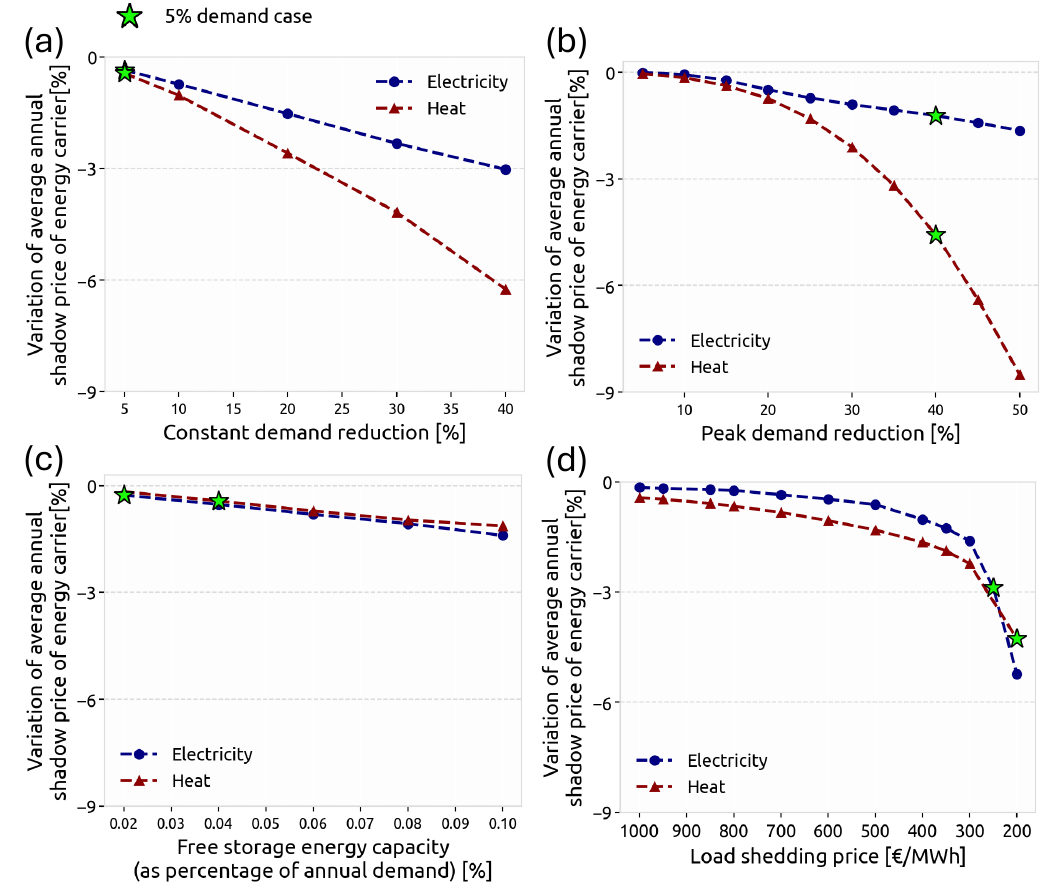}
\caption{\fontsize{9}{10}\selectfont Changes in the annual demand-weighted mean of shadow price of electricity and heating for (a) constant demand reduction, (b) peak shaving, (c) demand shifting with free storage, and (d) load shedding. The green stars (see Figure \ref{fig:3} of main text) represent scenarios altering roughly 5\% of annual demand. Average nodal marginal price of electricity and heating for the base scenario can be seen in Figures \ref{fig:S12}-\ref{fig:S13}.
}
\label{fig:S14}
\end{figure}

\newpage
\begin{figure}[H]
\renewcommand*{\thefigure}{S\arabic{figure}} \renewcommand{\figurename}{Figure}  

\includegraphics[width=0.7\textwidth, center]{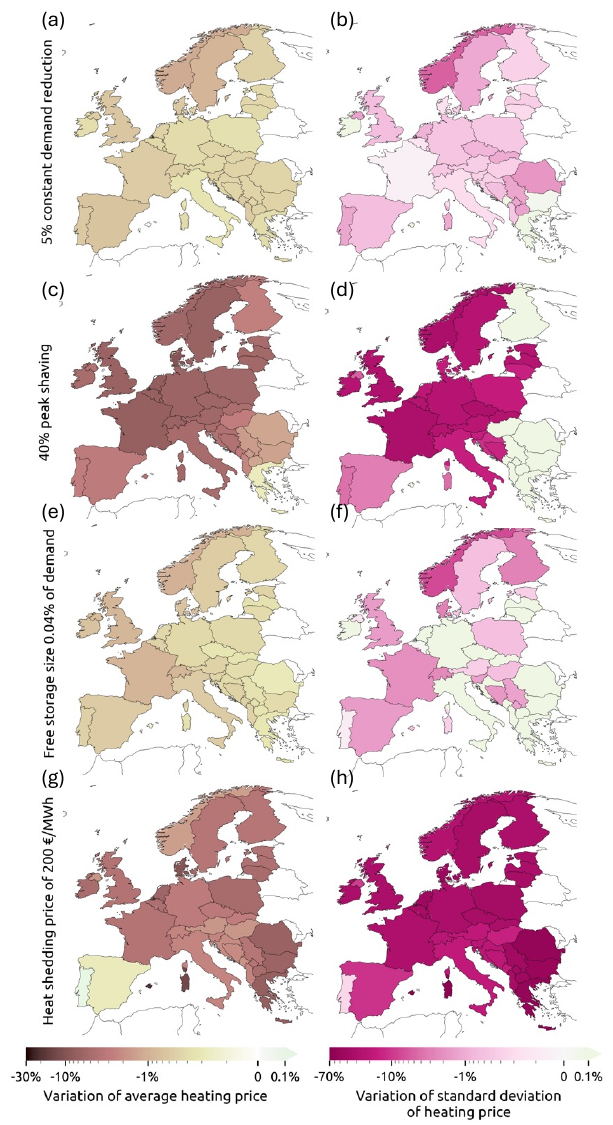}
\caption{\fontsize{8}{10}\selectfont Country-wise changes in shadow price of heating, which can be thought of as what consumers pay for heating, for (a) 5\% annual demand reduction, (c) 40\% peak shaving, (e) free storage size equal to 0.04\% of annual heating demand, and (g) heat shedding with price of 200 \euro/MWh. Figures (b), (d), (f), and (h) show the changes in standard deviation of heating price, which can be thought of as price volatility, for the mentioned scenarios, respectively.  The selected scenarios are almost equivalent in terms of annual heating demand that is altered (see green star scenarios in Figure \ref{fig:S14}).}
\label{fig:S15}
\end{figure}

\newpage

\begin{figure}[H]
\renewcommand*{\thefigure}{S\arabic{figure}} \renewcommand{\figurename}{Figure}  
\includegraphics[width=0.95\textwidth, center]{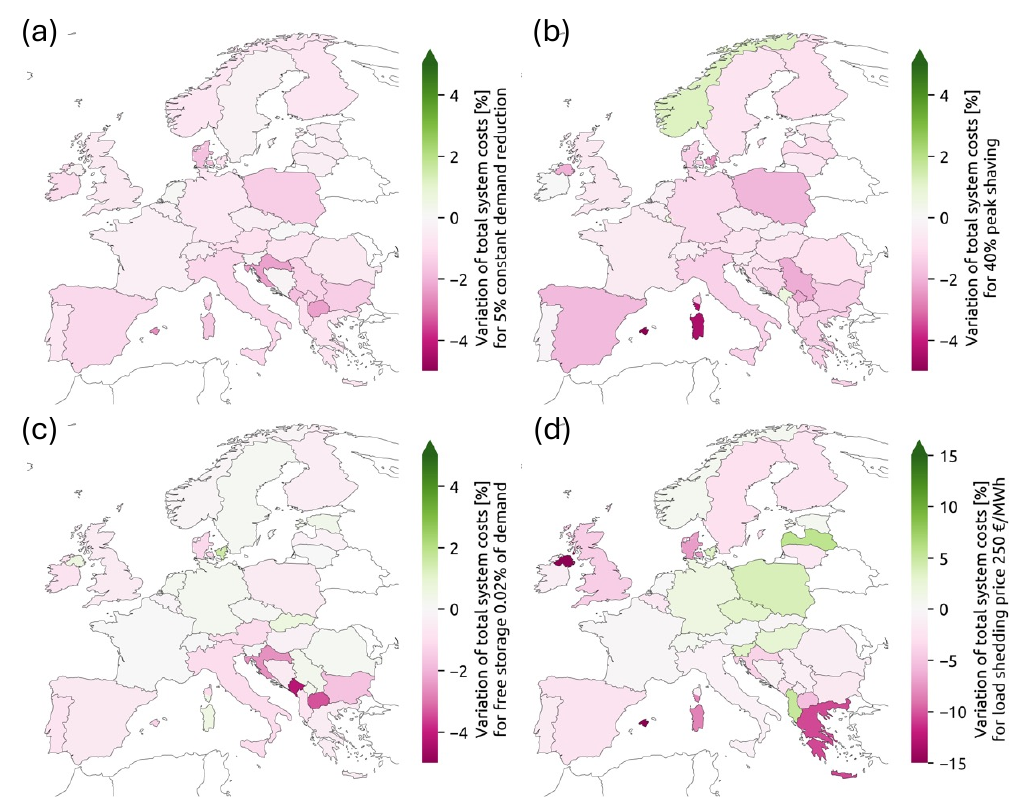}
\caption{\fontsize{9}{10}\selectfont Changes in total system cost for each country (sum of capital costs and marginal costs, excluding pipelines and transmission lines) under scenarios that are roughly equivalent in terms of annual electricity demand being altered (see green star scenarios in Figure \ref{fig:S14}): (a) 5\% annual demand reduction, (b) 40\% peak shaving, (c) free storage size equal to 0.02\% of annual demand, and (d) load shedding with price of 200 \euro/MWh. Note that while total system cost decrease for all scenarios, system costs for individual countries can increase }
\label{fig:S16}
\end{figure}

\begin{figure}[!htbp]
\renewcommand*{\thefigure}{S\arabic{figure}} \renewcommand{\figurename}{Figure}  
\includegraphics[width=0.95\textwidth, center]{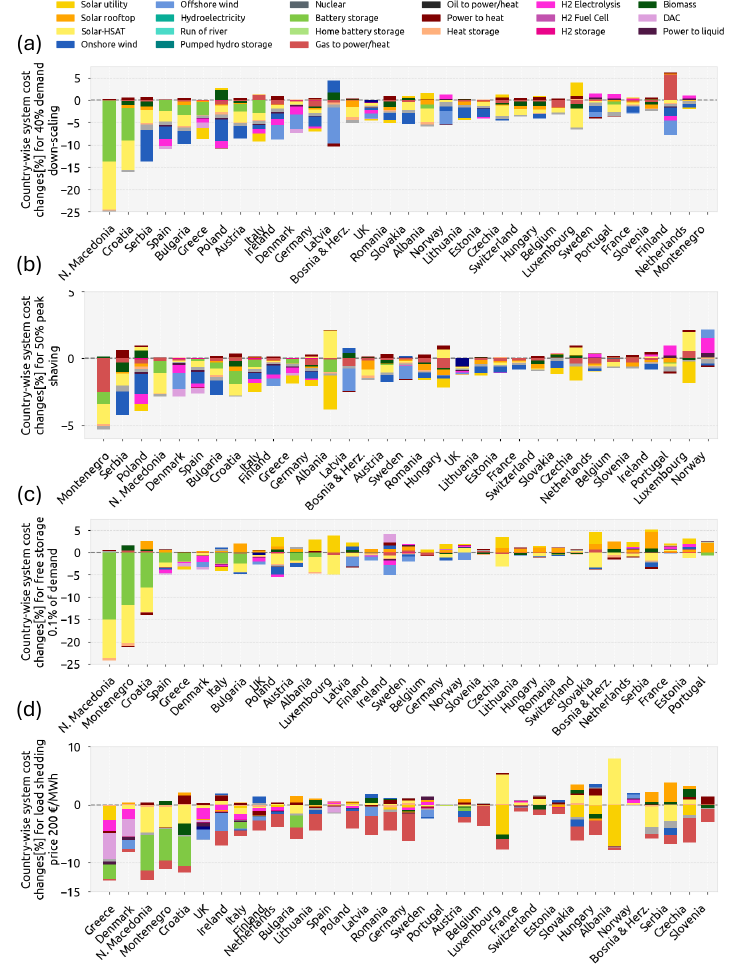}
\caption{\fontsize{8}{10}\selectfont Share of each technology in country-wise system cost changes under different demand mechanisms for the electricity sector: (a) 40\% annual demand reduction, (b) 50\% peak shaving, (c) free storage size equal to 0.1\% of annual demand, and (d) load shedding with price of 200 \euro/MWh. For example, when down-scaling electricity demand by 40\% in (a), total system cost for North Macedonia is reduced by 25\%, and almost half of this cost reduction is achieved by reducing the cost of battery in the country. Total system cost for each country is the sum of capital costs and marginal costs, excluding pipelines and transmission lines. Countries are ordered by the net reduction in total system cost, meaning the left-most country in each figure has the highest system cost reduction, while the right-most country has the lowest reduction or the highest increase.
}
\label{fig:S17}
\end{figure}

\newpage

\end{spacing}

%\clearpage
%\newpage
%\fontsize{9}{10}\selectfont 
%\addcontentsline{toc}{section}{Supplementary References}
%\bibliographystyleS{elsarticle-num}
%\bibliographyS{Bibliography.bib}
%\bibliographystyle{elsarticle-num}
%\bibliographystyleS{naturemag}
%\bibliography{Bibliography.bib}

%\restoregeometry

%\newpage
%\fontsize{9}{10}\selectfont
%\addcontentsline{toc}{section}{Supplementary References}
%\bibliography{Bibliography.bib}

%\end{document}

%\end{spacing}

%\clearpage
%\tableofcontents

\end{document}